\documentclass[12pt]{article}
\usepackage{amsmath,amsfonts,amssymb}
\usepackage{tabu}
\usepackage{amstext}
\usepackage{graphicx}
\usepackage{dcolumn}
\usepackage{bm}
\usepackage{graphics}
\usepackage{subfigure}
\usepackage{color}
\usepackage{cite}
\def\ub{\bar{u}}
\def\db{\bar{d}}

\def\ubcs{\bar{u}^{cs}}
\def\dbcs{\bar{d}^{cs}}

\def\sds{s^{ds}}

\usepackage{color}
\definecolor{gray}{rgb}{0.6,0.6,0.6}
\definecolor{darkgreen}{rgb}{0.0, 0.545098, 0.0}
\definecolor{darkblue}{rgb}{0.0, 0.0, 0.545098}

\usepackage[normalem]{ulem}

\begin{document}
\rightline{
	{\vbox {			
			\hbox{\bf MSUHEP-22-019}
}}}
\vspace*{0.2cm}

\begin{center}

{\bf{\large Connected and Disconnected Sea Partons from CT18 Parametrization of PDFs}}

\vspace{0.6cm}

{\bf  Tie-Jiun Hou\textsuperscript{1,*}, Mengshi Yan\textsuperscript{2}, Jian Liang\textsuperscript{3}, Keh-Fei Liu\textsuperscript{4}, C.-P.  Yuan\textsuperscript{5}}

\end{center}

\begin{center}
{\it
	{\bf 1} School of Nuclear Science and Technology, University of South China, Hengyang, Hunan 421001, China
	\\
	{\bf 2} Department of Physics and State Key Laboratory of Nuclear Physics and Technology, Peking University, Beijing 100871, China
	\\
	{\bf 3} Guangdong Provincial Key Laboratory of Nuclear Science, Institute of Quantum Matter, South China Normal University, Guangzhou 510006, China \\
	Guangdong-Hong Kong Joint Laboratory of Quantum Matter, Southern Nuclear Science Computing Center, South China Normal University, Guangzhou 510006, China 
	\\
	{\bf 4} Department of Physics and Astronomy, University of Kentucky, Lexington, KY 40506, U.S.A.
	\\
	{\bf 5} Department of Physics and Astronomy,
	Michigan State University, East Lansing, MI 48824, U.S.A.
	\\
	* tjhou@msu.edu
}
\end{center}

\begin{abstract}
The separation of the connected and disconnected sea partons, which were uncovered in the Euclidean path-integral formulation of the hadronic tensor, is accommodated with an extended parametrization of the non-perturbative parton distribution functions in the CT18 global analysis. This is achieved with the help of the distinct small $x$ behaviours of these two sea partons and the constraint from the lattice calculation of the ratio of the strange momentum fraction to that of the $\bar u$ or $\bar d$ in the disconnected insertion. The whole dataset of CT18 is used in this CT18CS fit. The impact of the recent SeaQuest data on the $\bar{d}(x)-\bar{u}(x)$ distribution of CT18CS is also discussed. The separate momentum fractions for the valence, the connected sea and disconnected sea of $u$ and $d$, the strange and the gluon partons are presented at the input scale $\mu =1.3$ GeV for the first time. They can be compared term-by-term with systematic error controlled lattice calculations.  
\end{abstract}
\begin{center}
\today
\end{center}

\newpage

\section{Introduction} \label{intro}

In high energy experiments, such as those at hadron colliders, theoretical analyses depend on the
parton structure of the hadronic beams in terms of their parton distribution functions (PDFs) in order
to understand the $W^{\pm}, Z$ and Higgs productions in precision measurements of the Standard Model parameters and the search of new physics. The universal PDFs can be extracted from deep inelastic 
scattering (DIS) and Drell-Yan processes with the help of factorization theorem and global analyses which involve the DGLAP evolution equations. Since the factorization formula involves an integral of the product of the parton distribution functions (PDFs) and the perturbative short distance kernel, extracting PDFs is intrinsically an inverse problem. The common approach is to
model the PDFs in terms of the valence and sea partons with respective small and large $x$ behaviours
and perform a global fit of the available experimental data at different $Q^2$ values. As a result, 
the quality of the fit and its accuracy depend on the precision and availability of the experimental data
in the relevant kinematic range. In particular, the flavour structure of the partons can be  improved with experiments which directly address the flavour dependence. For example, the first experimental evidence that the sea patrons have non-trivial flavour dependence is shown in the experimental demonstration of the violation of Gottfried sum rule. The original 
Gottfried sum rule, $I_G \equiv \int^1_0 dx [F^p_2(x)-F^n_2(x)]/x  =1/3$, was obtained
under the assumption that $\bar u$ and $\bar d$ sea partons are the same~\cite{Gottfried:1967kk}. However, the NMC measurement~\cite{NewMuon:1991hlj,NewMuon:1993oys}
of $\int^1_0 dx [F^p_2(x)-F^n_2(x)]/x$ turns out to be $0.235 \pm 0.026$, a 4 $\sigma$ difference from the Gottfried sum rule, which implies that the $\bar u = \bar d$ assumption was invalid. 
The recent SeaQuest experiment clearly shows that the $\bar{d}/\bar{u}$ ratio
in the range $0.1<x<0.4$ is substantially larger than unity ($\sim 1.5 $)~\cite{SeaQuest:2021zxb}.
Other flavour-dependent issues under active experimental and theoretical pursuits include the intrinsic strange and charm partons~
\cite{Brodsky:1980pb,Hou:2017khm,Ball:2016neh,NNPDF:2017mvq,Ball:2021leu}, and the 
$s(x) - \bar{s}(x)$~\cite{Davidson:2001ji,Kretzer:2003wy,Lai:2007dq,NNPDF:2017mvq,Ball:2021leu, Bailey:2020ooq} and $c(x) - \bar{c}(x)$~\cite{Sufian:2020coz} differences. 

The violation of the Gottfried sum rule prompted the Euclidean path-integral formulation of the 
the hadronic tensor of the nucleon for DIS which uncovered that there are two kinds of sea partons, one is the
connected sea and the other disconnected sea~\cite{Liu:1993cv,Liu:1999ak}. They are so named to reflect the topology of the quark lines in the 4-point current-current correlator in the nucleon. The connected sea (CS) results from a connected insertion of the currents on the same `valence' quark line and the disconnected sea (DS) is from a disconnected insertion involving a vacuum polarization from the quark loop involving  the external currents. These are `hand-bag' diagrams. On the other hand, the `cat's ears" diagrams, where two currents in the current-current correlator couple to different quark lines, are higher twists and are suppressed in the DIS region, but they are as important as the leading twists in low-energy lepton-nucleon scattering~\cite{Liang:2019frk,Liang:2020sqi}. The suppression of the higher twist contributions at large $Q^2$ has been demonstrated in a recent lattice calculation which shows that the ‘cat’s ears’ diagrams drops out quickly as compared to those ‘hand-bag’ diagrams when the three momentum transfer becomes large~\cite{Liang:2020sqi}. It is proved~\cite{Liu:1993cv} that, in the isospin symmetric limit, the Gottfried sum rule violation originates only from the CS which is subject to Pauli blocking due to the unequal numbers of the valence $u$ and $d$ quarks in both the proton and the neutron.  Attempts have been made~\cite{Liu:2012ch,Peng:2014uea,Peng:2014eza,Liang:2019xdx} to separate out the CS and DS partons by combining strange parton distribution from a HERMES experiment~\cite{Airapetian:2008qf}, $\bar{u} + \bar{d}$ from the CT10 analysis~\cite{Lai:2010vv}, and the ratio $\langle x\rangle_{s} / \langle x\rangle_ u ({\rm disconnected\,\, insertion})$ from lattice calculations~\cite{Doi:2008hp,Liang:2019xdx}. 

In this work, we shall accommodate parton degrees of freedom delineated in the path-integral formulation of the hadronic tensor in the form of CT18 global analysis \cite{Hou:2019efy} of unpolarized PDFs. Adopting lattice
results as constraints to perform the global fits has been applied to quark transversity distribution~\cite{Lin:2017stx}. The present work goes one step further to explicitly separate the CS and DS degrees of freedom for the first time under the CT18 parametrization~\cite{Hou:2021uoj}. 

This manuscript is organized as follows. Sec.~\ref{HTparton} gives a brief review of the path-integral formulation of the hadronic tensor which defines the parton degrees of freedom. Sec.~\ref{fitting} describes the parametrization of the PDF for each of the parton degrees of freedom and the details
of the global analysis. The result of a global analysis with the inclusion of both CS and DS partons fitted to the original CT18 data sets, termed as CT18CS fit, is presented in Sec.~\ref{results}. The second moments of the separate valence, the disconnected sea, and the gluon partons are presented for the first time at the input scale which can finally be compared directly with lattice calculations for each term and each flavor.
We note that the E906 SeaQuest \cite{SeaQuest:2021zxb} data only became available after the completion of the CT18 analysis.
Hence, we shall examine in Sec.~\ref{E866E906} the impact of the E906 SeaQuest \cite{SeaQuest:2021zxb} data on a global fit similar to CT18CS, in which the E866 NuSea \cite{NuSea:2001idv} data was already included.
Sec.~\ref{summary} contains our summary. 

\section{Parton Degrees of Freedom from Euclidean Path-integral Formulation of the Hadronic Tensor}  \label{HTparton}

The Euclidean hadronic tensor was formulated in the path-integral formalism to identify the origin of the Gottfried sum rule violation~\cite{Liu:1993cv,Liu:1999ak}. It is defined as the current-current correlator in the nucleon 
with Fourier transform in the spatial directions 
\begin{eqnarray}  \label{wmunu_tilde}
\widetilde{W}_{\mu\nu}(\vec{q},\vec{p},\tau)
    &=& <N(\vec{p})| \int d^3x  \frac{e^{i \vec{q}\cdot \vec{x}}}{4\pi} 
J_{\mu}(\vec{x},\tau) J_{\nu}(0,0)|N(\vec{p})>,
\end{eqnarray}
It is a function of $\tau$, which is the Euclidean time separation between the currents.
Formally, the inverse Laplace transform converts $\widetilde{W}_{\mu\nu}(\vec{q},\vec{p},\tau)$ to the Minkowski hadronic tensor
\begin{equation}  \label{wmunu}  
W_{\mu\nu}(\vec{q},\vec{p},\nu) = \frac{1}{2m_Ni} \int_{c-i \infty}^{c+i \infty} d\tau\,
e^{\nu\tau} \widetilde{W}_{\mu\nu}(\vec{q},\vec{p}, \tau),
\end{equation} 
with $c > 0$. However, this is not practical in lattice calculation, as there are no data on the imaginary 
$\tau$. Instead, one can turn this into an inverse problem and find a solution from the Laplace transform~\cite{Liu:2016djw}
  \begin{equation}  \label{Laplace}
 \widetilde{W}_{\mu\nu} (\vec{q},\vec{p},\tau) = \int d \nu\, e^{-\nu\tau}
 W_{\mu\nu}(\vec{q},\vec{p},\nu).
  \end{equation}
  This has been studied~\cite{Liu:2016djw,Liang:2017mye,Liang:2019frk,Liang:2020sqi} with the inverse algorithms such as Backus-Gilbert,  Maximum Entropy  and Bayesian Reconstruction methods. 
The spectral density in lepton-nucleon scattering has several kinematic regions as the energy transfer 
$\nu$ increases -- the elastic scattering, the inelastic reactions ($\pi N, \pi\pi M, \eta N$ etc.) and resonances ($\Delta$, Roper, $S_{11}$, etc.), shallow inelastic scattering (SIS), and deep inelastic scattering  (DIS) regions. To determine how large a $\nu$ is needed for DIS, we look at $W$, the total invariant mass of the hadronic state for the nucleon target at rest
\begin{equation}
W^2 = (q+p)^2 = m_p^2 - Q^2 + 2 m_p \nu
\end{equation}
 The global analyses of the high energy lepton-nucleon and Drell-Yan experiments to extract the parton distribution functions usually make a cut with $W^2 > 12\, {\rm GeV^2}$ to avoid the
 elastic and inelastic regions. Thus, to be qualified in the DIS region, the energy transfer  $\nu$ needs to be greater than 8 GeV for $Q^2 = 4\, {\rm GeV^2}$, a typical choice made in the CTEQ-TEA PDF global analysis. 
     
\begin{figure}[htpb]
\centering
\subfigure[]
{{\includegraphics[width=0.32\hsize]{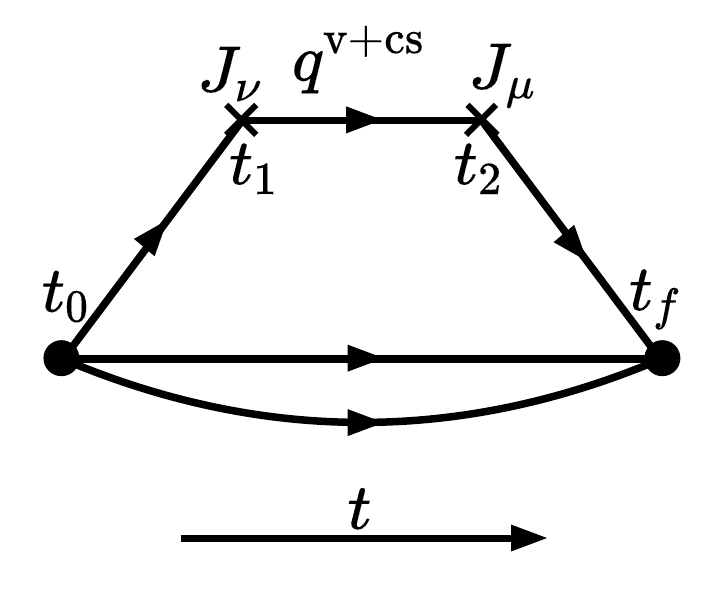}}
  \label{v+CS}}
\subfigure[]
{{\includegraphics[width=0.32\hsize]{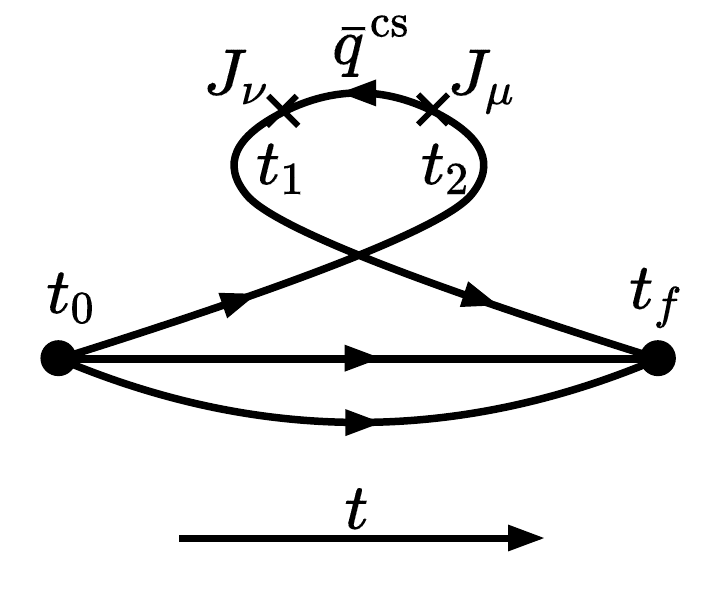}}
  \label{CS}}
\subfigure[]
{{\includegraphics[width=0.32\hsize]{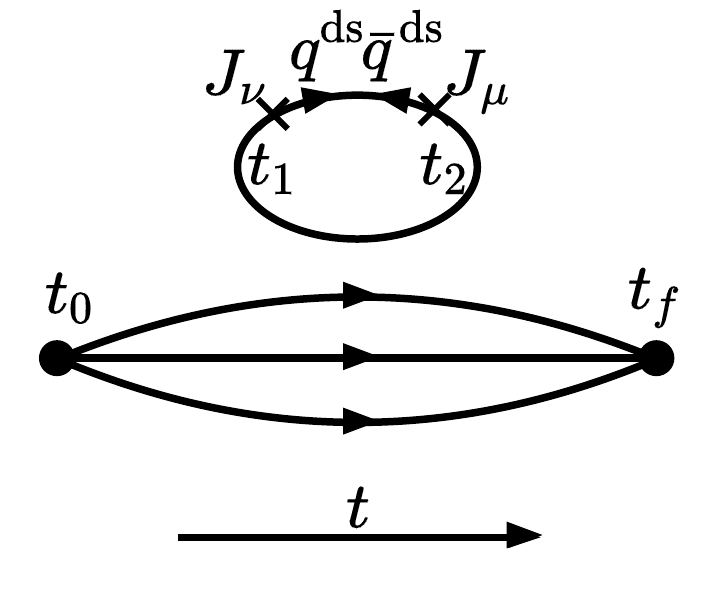}}
 \label{DS}}
 \caption{Three gauge invariant and topologically distinct insertions in the Euclidean-path integral
formulation of the nucleon hadronic tensor where the currents couple to the same quark
propagator. In the DIS region, the parton degrees of freedom are
  (a) the valence and connected sea (CS) quarks $q^{v+cs}$, (b) the CS anti-quarks $\bar{q}^{cs}$. Only $u$ and $d$ are present in (a) and (b) for the nucleon hadronic tensor. (c) the disconnected sea (DS) quarks $q^{ds}$ and anti-quarks $\bar{q}^{ds}$ with $q = u, d, s,$ and $c$.\label{leading-twist}} \vspace*{1cm}
 \subfigure[]
{{\includegraphics[width=0.32\hsize]{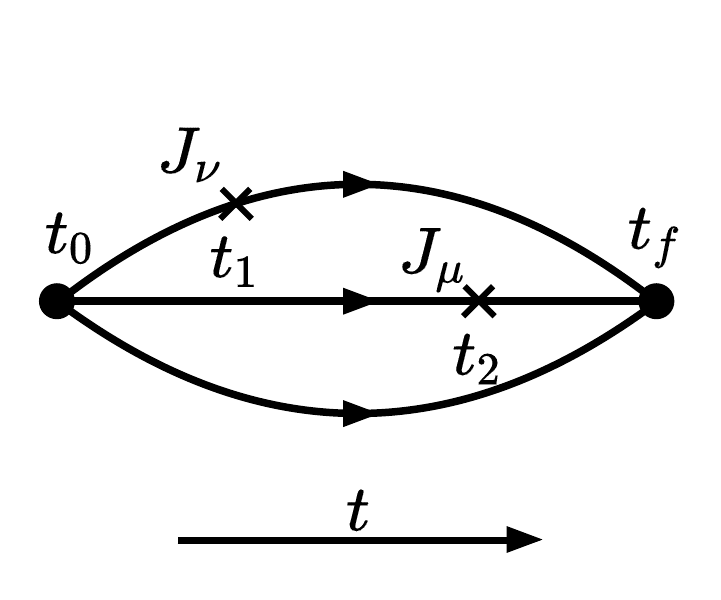}}
  \label{cat_ear_1}}
\subfigure[]
{{\includegraphics[width=0.32\hsize]{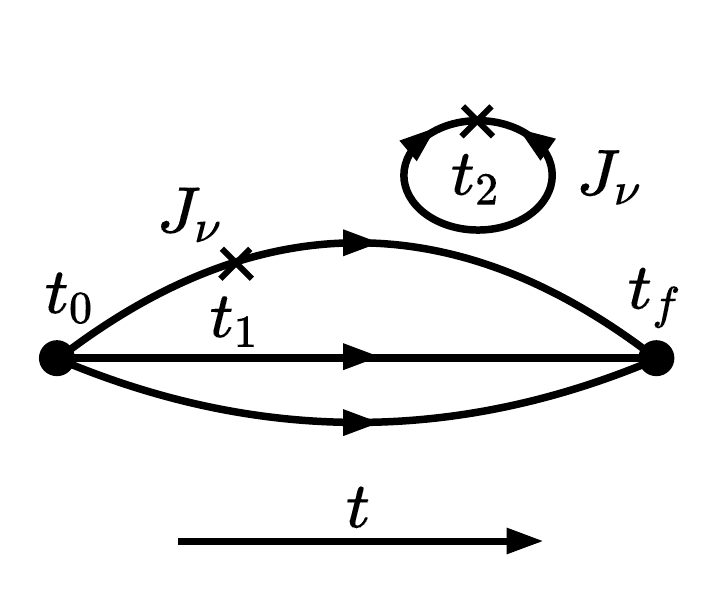}}
  \label{cat_ear_2}}
\subfigure[]
{{\includegraphics[width=0.32\hsize]{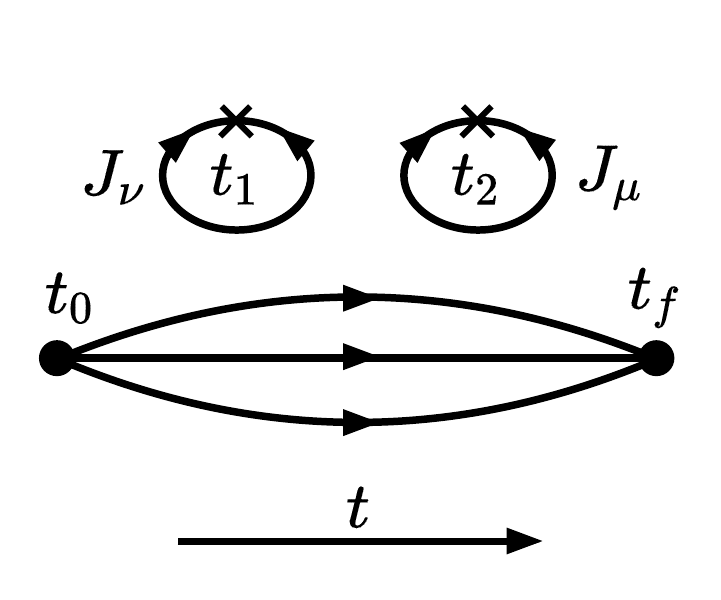}}
 \label{cat_ear_3}}
\caption{Three other gauge invariant and topologically distinct insertions where the currents are
inserted on different quark propagators. In the DIS region, they are higher-twist diagrams.
\label{higher-twist}}
\end{figure}

 It is shown~\cite{Liu:1993cv,Liu:1999ak,Liu:1998um,Liu:2020okp} that, when the time ordering 
 $t_f > t_2 > t_1 > t_0$ is fixed, the 4-point function for extracting the matrix element 
 $\widetilde{W}_{\mu\nu}(\vec{q},\vec{p},\tau)$ in Eq.~(\ref{wmunu_tilde})
 can be grouped in terms of 6 topologically distinct and gauge invariant path-integral insertions, according to different Wick contractions among the Grassmann numbers in the two currents and the source/sink interpolation fields. They can be further grouped into two classes. The first class includes those where the
currents are coupled to the same quark propagator as illustrated in Fig.~\ref{leading-twist}. The second class involves those where the two currents are coupled to different quark propagators as illustrated in Fig.~\ref{higher-twist}. In low energy lepton-nucleon scattering, all 6 diagrams contribute and they are not separable~\cite{Liu:2020okp}. However, in the DIS region,
the first class are `hand-bag' diagrams which include the leading twist contributions, the second class are `cat's ears' diagrams, which are higher twists and are suppressed by $\mathcal{O}(1/Q^2)$.

The first class in Fig~\ref{leading-twist} includes three path-integral
diagrams that can be denoted as connected insertions  (CI) (Fig.~\ref{v+CS} and Fig.~\ref{CS}), where the quark lines are all connected, and disconnected insertions (DI) (Fig.~\ref{DS}), where there are vacuum polarizations associated with the currents in disconnected quark loops.
We should note that Fig.~\ref{CS} includes the exchange contribution to prevent the $u$ or $d$ quark in the loop in Fig.~\ref{DS} from occupying the same Dirac eigenstate in the nucleon propagator, enforcing the Pauli principle. In fact, Fig.~\ref{DS} and Fig.~\ref{CS} are analogous to the direct and exchange diagrams in time-ordered Bethe-Goldstone diagrams in the many-body theory~\cite{Bethe:1957}.

As far as the leading-twist DIS structure functions $F_1, F_2$ and $F_3$ are concerned,  the three diagrams in Fig.~\ref{leading-twist} are additive with contributions classified as the valence and sea quarks $q^{v+cs}$ in Fig.~\ref{v+CS}, the connected sea (CS) antiquarks $\bar{q}^{cs}$ in Fig.~\ref{CS}, 
and disconnected sea (DS) quarks $q^{ds}$ and antiquarks $\bar{q}^{ds}$ in Fig.~\ref{DS}~\cite{Liu:1993cv,Liu:1999ak,Liu:1998um,Liu:2020okp}. Since the $u$ and $d$ partons in the quark loop in Fig.~\ref{DS} appear in a different flavour trace than the one involving the nucleon propagator, they cancel in the Gottfried sum in the isospin symmetric limit. Thus, the Gottfried sum rule violation comes entirely from the connected sea (CS) difference $\bar{u}^{cs} - \bar{d}^{cs}$ in the $F_2$ structure functions in this case~\cite{Liu:1993cv}. It is proven~\cite{Liu:1999ak,Liu:2020okp} from short distance expansion that the parton degrees of freedom defined in diagram in Fig.~\ref{leading-twist} are
separable, unlike the case of low-energy lepton-nucleon scattering, where the higher twists are important~\cite{Liang:2019frk,Liang:2020sqi}. Furthermore, these parton degrees of freedom are identical to those defined from the recent Feynman-$x$ approaches~\cite{Liu:2020okp}, i.e. quasi-PDF~\cite{Ji:2013dva}, pseudo-PDF~\cite{Radyushkin:2017cyf}, and lattice cross section~\cite{Ma:2017pxb}. 

PDFs can be extracted from the factorization formula~\cite{Collins:1989gx} where the experimental cross section or structure functions are expressed as a convolution integral of the coefficient functions and
the PDFs. In practice, the global fitting programs adopt the parton degree of freedoms as $u, d, \bar{u}, \bar{d}, s, \bar{s}$ and $g$. We see that from the path-integral formalism of QCD, each of the
$u$ and $d$ have two sources, one from the connected insertion (CI) (Fig.~\ref{v+CS}) and one
from the disconnected insertion (DI) (Fig.~\ref{DS}), so are $\bar{u}$ and $\bar{d}$ from Fig.~\ref{CS}
and Fig.~\ref{DS}. On the other hand, $s,c$ and $\bar{s},\bar{c}$ only come from the DI (Fig.~\ref{DS}).
In other words,
\begin{align}  \label{dof}
      &u=u^{v+cs} + u^{ds},             &   & d= d^{v+cs} + d^{ds}, \nonumber \\
&\bar{u}=\bar{u}^{cs} + \bar{u}^{ds},   &   & \bar{d}= \bar{d}^{cs} + \bar{d}^{ds}, \nonumber \\
      &s=s^{ds},                        &   & \bar{s}= \bar{s}^{ds},     \nonumber \\
      &c=c^{ds},                        &   & \bar{c}= \bar{c}^{ds}.
\end{align}

This classification of the parton degrees of freedom is richer than those in terms of $q$ and
 $\bar{q}$ in the global analysis due to the fact that there are two sources for the quarks -- $q^{v+cs}$ and $q^{ds}$ -- and two sources for the antiquarks -- $\bar{q}^{cs}$ and $\bar{q}^{ds}$. The distinguishing
feature of CS and DS lies in their characteristic small-$x$ behaviors, which we shall explore in this work
to perform global analysis. In the Regge theory, the small-$x$ behaviour of $q^{v+cs}$ and $\bar{q}^{cs}$,
being in the flavour non-singlet connected insertions, are dominated by the reggeon exchange. Thus, we expect  $q^{v+cs}(x), \bar{q}^{cs}(x) {}_{\stackrel{\longrightarrow}{x \rightarrow 0}} \, x^{-\alpha}$ for $q = u, d$, where  $\alpha \sim 0.5$ is  the slope of the Regge trajectory. 
Whereas, DS is flavour
singlet and can have Pomeron exchanges. Hence, $q^{ds}(x), \bar{q}^{ds} {}_{\stackrel{\longrightarrow}{x \rightarrow 0}} \, x^{-1}$ for $q = u,d,s,c$. In an attempt to separate the CS
and DS quarks~\cite{Liu:2012ch} by combining strange quark distribution from a HERMES experiment
and $\bar{u} + \bar{d}$ from CT10, it is found that $x(\bar{u}^{cs} - \bar{d}^{cs})$ spans the same $x$ range as that of $x(\bar{u} - \bar{d})$, which suggests that they have similar small-$x$ behaviours;
whereas $x(\bar{u}^{ds} + \bar{d}^{ds})$ is much singular for $x < 0.05$~\cite{Liu:2012ch}. This is consistent with expectation. 

Until the Feynman-$x$  and/or the hadronic tensor approaches on the lattice have all the systematic
errors, such as excited states contamination and large nucleon momentum, are under control so 
that all region of $x$ can be compared with those from the global analyses, the most reliable comparison
between global fittings and lattice calculations are via the parton moments. The latter are getting
mature with all the systematic errors (e.g. continuum and infinite volume extrapolations, excited states
contamination, physical pion mass, non-perturbative renormalization, and scale setting) have been
taken into account~\cite{Lin:2017snn,Lin:2020rut}. However, as pointed out in~\cite{Liu:2017lpe,Liu:2020okp}, it is not possible to compare the moments
from global analyses and those from the lattice calculations in detail, except for the limited isovector 
($u-d$) and stangeness moments. This is because the lattice calculation of moments in the three-point functions are organized in the connected insertions (CI) and disconnected insertions (DI). The CI includes
both $q^{v+cs}$ and $\bar{q}^{cs}$, while DI includes $q^{ds}$ and $\bar{q}^{ds}$. On the other hand,
in the present global analyses, CS and DS degrees of freedom are not separated. To make a comparison at the moment level, it is encumbered upon global analyses to disentangle the connected sea from the disconnected, so that the full lattice results of moments in CI and DI can be compared to them for each flavor.

\section{Global fitting} \label{fitting}

In this section, the general setting of the CT18CS global fit is presented. The CT18CS, as an extended parametrisation of PDFs in accommodation with the Euclidean path-integral formalism of QCD, requires a different scheme of parton classification with more parton degrees of freedom. 
The specific parton degrees of freedom to be parametrized and a number of ansatzes imposed in this global analysis will be explained in the following Sec.~\ref{sub:PartonDoF}.
In the Sec~\ref{sub:Small-Largex}, we will introduce the settings of small-$x$ and large-$x$ behaviour for CS and DS parton distributions.

\subsection{Parton degrees of freedom} \label{sub:PartonDoF}

In the QCD global analysis of parton distributions in the proton, the PDFs of various partons are parametrized in some functional forms at the initial scale $Q_0$ (about 1 GeV), from where the PDFs are evolved to any arbitrary higher energy scale $Q$ via DGLAP evolution equations. 
Typically, it is assumed that the charm and bottom quark PDFs are generated perturbatively from QCD evolution, though in some special studies, the possibility of having non-perturbaive charm PDF at the $Q_0$ scale was also considered, such as in Refs.~\cite{Hou:2017khm,Ball:2016neh,NNPDF:2017mvq,Ball:2021leu, Ball:2022qks}. 
Therefore, in general, 
the total number of parton degrees of freedom at the $Q_0$ scale is 7, which includes the following partons: 
\begin{equation} \label{c_dof}
g, \, u^v, \, d^v, \, \bar{u}, \, \bar{d}, \, s, \, \bar{s} 
\end{equation}
In CT18~\cite{Hou:2019efy}, it is also assumed that the strange PDFs $s = \bar{s}$ at the $Q_0$ scale, though $s \neq \bar{s}$  will be generated at large $Q$ scale via NNLO QCD evolution. Given this ansatz, the number of parton degrees of freedom is 6 in CT18. 
We note that in MSHT20 PDFs~\cite{Bailey:2020ooq} and NNPDF4.0~\cite{Ball:2021leu}, a non-vanishing asymmetric strangeness $s(x) \neq \bar{s}(x)$ is imposed in the non-perturbative parametrisation at their respective $Q_0$ scales.

As mentioned in the last section, when the separation of CS partons and DS partons are considered, we would have more partonic degrees of freedom. The classification of partons becomes: 
\begin{equation} \label{e_dof}
g, \, u^{v+cs}, \, u^{ds}, \, \bar{u}^{cs}, \, \bar{u}^{ds}, \, d^{v+cs}, \, d^{ds}, \, \bar{d}^{cs}, \, \bar{d}^{ds}, \, s^{ds}, \, \bar{s}^{ds},
\end{equation}
totally 11 of them.
To implement all the degrees of freedom in Eq.~(\ref{e_dof}) and obtain their $Q^2$ dependence would require generalized DGLAP evolution equations as developed in Ref.~\cite{Liu:2017lpe}. In the present study, we shall pararmetrize the extended set of partons in Eq.~(\ref{e_dof}) at the input scale $Q_0$. with some specific assumptions to be listed below, and then combine the CS and DS into the 
conventional partons in Eqs.~(\ref{dof}) and (\ref{c_dof}) so that we can use the same NNLO evolution equations as for CT18. In this way, we can
compare with the results of CT18 to discern the different roles played by CS and DS and their respective impacts on physics at this stage. When the generalized evolution code is ready, we can fully explore the CS and DS effects at all scales.

For the present work, we have adopted the following assumptions to  
reduce the number of parton degrees of freedom  from 11 to 6, similar to that in CT18.

\begin{itemize}
    \item Similar to CT18, we assume the symmetric disconnected sea parton distributions:
\end{itemize}
\begin{equation}
    u^{ds} = \bar{u}^{ds}, d^{ds} = \bar{d}^{ds}, \text{ and } s^{ds} = \bar{s}^{ds}.
\end{equation}
\begin{itemize}
    \item The isospin symmetry is imposed for the $u$ and $d$ quarks:
\end{itemize}
\begin{equation}
u^{ds} = \bar{u}^{ds} = d^{ds} = \bar{d}^{ds}.
\label{eq:ansatz_equation}
\end{equation}
\begin{itemize}
    \item The DS components of $u$ and $d$ quark PDFs are proportional to the $s$ quark PDF, i.e.

\begin{equation}
    u^{ds} = \bar{u}^{ds} = d^{ds} = \bar{d}^{ds} = R \times s.
\label{eq:ds_strange}
\end{equation}
Since the DS in the lattice calculation involves a correlation between the quark loop and the nucleon propagator via the gluons, it is not as sensitive to the nucleon wavefunction as are the valence and CS partons in the connected insertion. The only difference between $u^{ds}$,  $d^{ds}$ and $s$ is their quark masses. Thus, it is reasonable to postulate that their distribution are the same modulo a proportional constant $R$.
In this work, we determine the value of $R$ from the ratio of the second moment between the strange and the sum of $u$ and $d$ in the disconnected insertion, predicted by a lattice QCD calculation which has taken all the systematic errors into account~\cite{Liang:2019xdx}. It yields $1/R = \langle x \rangle_{s+\bar{s}}/\langle x \rangle_{\bar{u}+\bar{d}}(\text{DI}) = 0.822(69)(78)$ at 1.3 GeV, where $\langle x \rangle_{\bar{u}+\bar{d}}(\text{DI})$ is the momentum fraction carried by the light quark (either $u$ or $d$) in the disconnect insertions. This result was obtained by properly evolving the  matching coefficients from 2 GeV to 1.3 GeV~\cite{Yang:2018nqn}, using the known result of $1/R$ at 2 GeV, which was found to be  $0.795(79)(77)$~\cite{Liang:2019xdx}. 

In the CTEQ-TEA PDF global analysis, the 
normalizations for individual sea quark PDFs are computed using the valence quark and momentum sum rules, and the first moments $\langle x \rangle_{g}$ and the ratio $\langle x\rangle_{\bar s+\bar s}/\langle x\rangle_{\bar u+\bar d}$ fitted as free parameters. Since the parametrizations do not determine the ratio of the strange-to-nonstrange PDFs, we restrict this ratio in the present work by the above-mentioned prediction from lattice-QCD. Specifically, we require that the ratio  $\left({s}^{ds}(x) + \bar {s}^{ds}(x)\right)/\left(\bar{u}^{ds}(x) + \bar{d}^{ds}(x)\right)$, at $Q_0=1.3$ GeV,  is 
constrained at the 68\% confident level to be in the interval $[0.718, 0.926]$ with a central value of 0.822,
 by imposing the appropriate Lagrange Multiplier constraint in the CT18CS fit.

Finally, we note that the assumption in Eq.~(\ref{eq:ds_strange}) can be checked by the similar $\langle x^3\rangle$ ratio in lattice calculations in the future.
\end{itemize}

\begin{itemize}
    \item We further define 
\end{itemize}
\begin{equation}
    u^{cs} \equiv \bar{u}^{cs}\text{ and }d^{cs} \equiv \bar{d}^{cs}, 
\label{eq:cs_ansatz}
\end{equation}
so that
\begin{equation} \label{valence}
	u^v = u^{v+cs} - u^{cs} = u^{v+cs} - \bar{u}^{cs}, 
\end{equation}
which agrees with the usual definition of the {\it valence} quark: $q^v \equiv  q - \bar q =[u^{v+cs} +u^{ds}] - [\bar{u}^{cs}+\bar{u}^{ds}] $,
when $u^{ds}=\bar{u}^{ds}$. It was pointed out in~\cite{Liu:2017lpe} that when $u^{ds}$ and $\bar{u}^{ds}$ are not equal, the $q^v \equiv  q - \bar q$ definition leads to conceptual puzzles, such as the valence $u$ can evolve into valence $d$ in NNLO evolution and that the strangeness can have valence distribution when $s\neq \bar{s}$. These puzzles are resolved with the 
definition in Eq.~(\ref{valence})~\cite{Liu:2017lpe}.

With all the above conditions taken into account, the remaining parton degrees of freedom are $g, u^v, u^{cs}, d^v, d^{cs}, s^{ds}$. 
As discussed earlier below Eq.~(\ref{e_dof}), we shall combine CS and DS into
the usual $\bar{u}/\bar{d}$ d.o.f., i.e. $\bar{u} = \bar{u}^{cs} + \bar{u}^{ds} = \bar{u}^{cs} + Rs$ and $\bar{d} =\bar{d}^{cs} + \bar{d}^{ds} = \bar{d}^{cs} + Rs$ at the input scale and evolve them in the same NNLO equations as CT18 in the global fitting.

\subsection{Small-$x$ and large-$x$ behaviour} \label{sub:Small-Largex}

At the starting $Q_0$ scale, the non-perturbative PDFs are parametrised as 
\begin{equation}
    f(x) = x^{a_1-1}(1-x)^{a_2-1}\text{Poly(x)},
\end{equation}
where the parameters $a_1$ and $a_2$ dominate the behaviour of PDFs as $x$ approaches 0 or 1, respectively, and the $\text{Poly}(x)$, constructed with a set of Bernstein polynomials in CTEQ-TEA PDF family, is responsible for the shape of PDFs in a wide range of $x$.
In practice, we implemented the following ansatzes to parametrize various parton distributions at the $Q_0$ scale.

\begin{itemize}
    \item $\bar{d}/\bar{u}\stackrel{x\rightarrow0}{\longrightarrow}$ 1.\\
    Based on the isospin symmetry in strong interaction, we require $\bar{u}$ and $\bar{d}$ to have the same small-$x$ behaviour, where the disconnected sea dominates. Specifically, this ansatz is implemented by setting $a_1^{\bar{u}}=a_1^{\bar{d}}$ to preserve the isospin symmetry in the small-$x$ region. This ansatz was also applied in the CT18 fit.
    See, Appendix C of Ref.~\cite{Hou:2019efy}
    
    \item $u^{ds}, \bar{u}^{ds}, d^{ds}, \bar{d}^{ds}, s^{ds}, \bar{s}^{ds} \stackrel{x\rightarrow0}{\longrightarrow}$ $x^{-1}$.\\
    Since the DS partons are flavour singlet and can have Pomeron exchanges, their small-$x$ behaviour goes like $x^{-1}$. Based on Eq.~(\ref{eq:ds_strange}), this ansatz is implemented by setting $a_1^{s}=0$, which is the value of the shape parameter $a_1$ of the strangeness PDF.
    We note that $a_1^{s}=0$ is consistent with the CT18 error PDF sets, with the value of $a_1^{s}$ of the CT18 central set shown in the first row of  Table~\ref{tab:CT18CS_a1a2}.

    \item $u^{cs}, d^{cs} \stackrel{x\rightarrow0}{\longrightarrow} u^v, d^v$. \\
    Like valence partons, the CS partons are in the connected insertion, which is flavour non-singlet.  Thus, we set the small-$x$ behaviour of CS partons to be the same as those of valence-quark partons: $a_1^{u^{cs}}=a_1^{d^{cs}}=a_1^{u^v}=a_1^{d^v}$.
    
    \item $d/u\stackrel{x\rightarrow1}{\longrightarrow} d/u$ of CT18. \\
    In the CT18 fit, the ratio $d/u$ was required to approach a finite number as $x\rightarrow1$.
    This assumption is also kept in the CT18CS fit, which is done by setting $a_2^{u^v} = a_2^{d^v} = a_{2,\text{CT18}}^{u^v} = a_{2,\text{CT18}}^{d^v}
    =3.036$.
Since the ratio $d/u$ at $x\rightarrow1$ is dominated by valence partons and the parameter $a_2$ controls the PDF behaviour as $x\rightarrow1$, we fix the $a_2$ values of valence partons as those in CT18 fit, for simplicity.  
    
    \item $\bar{d}/\bar{u}\stackrel{x\rightarrow1}{\longrightarrow} \bar{d}/\bar{u}$ of CT18. \\
    As shown in Refs.~\cite{Hou:2019efy,SeaQuest:2021zxb}, CT18 PDFs can describe reasonably well both the E866 NuSea~\cite{NuSea:2001idv} and E906 SeaQuest~\cite{SeaQuest:2021zxb}  data, though the SeaQuest data were not included in the CT18 fit, as they only became available after the completion of the CT18 fit. 
    Since both data sets provide important constraints on the ratio  $\bar{d}/\bar{u}$ as $x\rightarrow1$, and the CS component of anti-quarks dominates the sea parton behaviour in the large $x$ region, we set 
 $a_2^{\bar{u}^{cs}} = a_2^{\bar{d}^{cs}} = a_{2,\text{CT18}}^{\bar{u}} = a_{2,\text{CT18}}^{\bar{d}} = 7.737$
in the CT18CS fit, for simplicity.  

\end{itemize}

For a quick comparison, we list in  Table \ref{tab:CT18CS_a1a2} the 
fitted values of $a_1$ and 
$a_2$ parameters of various partons in the CT18 and CT18CS NNLO fits. The numbers marked with ``$\star$'' indicate that they are not fitted, but input values in the CT18CS fit. Note that we did not list the values of the other shape parameters used in these fits. In total, there are 28 such shape parameters to be fitted in both the CT18 and CT18CS fits, with the same number (6) of parton degrees of freedom. We note that the published CT18 PDF error sets include an additional pair of eigenvector sets to account for the larger error of gluon  PDF in the small-$x$ region.

\begin{table}[htp]
\begin{center}
\begin{tabular}{c|cccccc}
\hline
\\
{\bf CT18}    &  $u^v$       &    $d^v$         &    $g$         &   $\bar{u}$      &      $\bar{d}$     &    $s$         \\
\hline                                                                                     
$a_1$         &  $0.763$     &    $0.763$      &    $0.531$     &    $-0.022$      &      $-0.022$      &    $-0.022$    \\
$a_2$         &  $3.036$     &    $3.036$      &    $3.148$     &    $7.737 $      &      $7.737$       &    $10.31$     \\
\hline                                \\                                                     
{\bf CT18CS}  &  $u^{v}$       &    $d^{v}$        &    $g$         &   $\ubcs$       &      $\dbcs$        &    $\sds$      \\
\hline                                                                                     
$a_1$         &  $0.739$     &    $0.739$      &    $0.553$     &    $0.739$       &      $0.739$       &    $0.000\star$     \\
$a_2$         &  $3.036\star$     &    $3.036\star$      &    $3.371$     &    $7.737\star$       &      $7.737\star$       &    $11.57$     \\ 
\hline  
\end{tabular}
\end{center}
\caption{The fitted values of $a_1$ and 
	$a_2$ parameters of various partons in the CT18 and CT18CS NNLO fits. The numbers marked with ``$\star$'' indicate that they are not fitted, but input values in the CT18CS fit.	
	Furthermore, the ansatz made in the CT18CS fit includes
		$u^{cs}=\bar{u}^{cs}\text{ and }d^{cs}=\bar{d}^{cs}$, cf. Eq.~(\ref{eq:cs_ansatz}), 
		and
		$u^{ds} = \bar{u}^{ds} = d^{ds} = \bar{d}^{ds} = R \times s$, with $1/R = \langle x \rangle_{s+\bar{s}}/ \langle x \rangle_{\bar{u}+\bar{d}}(\text{DI}) = 0.822(69)(78)$ at 1.3 GeV,  cf. Eq.~(\ref{eq:ds_strange}).
}
\label{tab:CT18CS_a1a2}
\end{table}

 \section{Results} \label{results}

In this section, we present the results of the CT18CS global fit on aspects of quality of the fit, the configuration of PDFs, and various PDF Mellin moments. The comparison  between CT18CS and the standard CT18 NNLO fits shows that CT18CS, with an extended parametrisation, is consistent with the CT18 global analysis. Note that this global analysis uses the same data sets as the ones used in the CT18 analysis. There are in total 39 data sets, with 3681 data points included~\cite{Hou:2019efy}.

\subsection{Quality of the fit} \label{sub:quality}

In Table \ref{tb:chi2}, we compare the quality of the CT18CS fit to that of the CT18 fit. 
It turns out, both have the total $\chi^2_{\text{CT18}} = 4292$  for a total of 3681 data points. 
The experimental data sets which made non-negligible contributions to the change in $\chi^2$ of these two fits are also listed in Table \ref{tb:chi2} for comparison. As expected, they are the data most sensitive to valence and sea quark PDFs.

\begin{table}[htpb]
\begin{center}
\begin{tabular}{cl|r|rr}
ID & Experimental data set & $N_{pt, E}$ & $\chi^2_{\text{CT18}}$ & $\chi^2_{\text{CT18CS}}$ \\ \hline \hline
 104 & NMC $F_2^d/F_2^p$ \cite{NewMuon:1996fwh} & 123 & 125.7 & 120.6 \\
 124 & NuTeV $\nu\mu\mu$ SIDIS \cite{Mason:2006qa} & 38 & 18.49 & 19.75 \\
 125 & NuTeV $\bar\nu \mu\mu$ SIDIS \cite{Mason:2006qa} & 33 & 38.45 & 40.05 \\
 201 & E605 Drell-Yan process \cite{Moreno:1990sf} & 119 & 103.4 & 107.1 \\
 203 & E866 Drell-Yan process $\sigma_{pd}/(2\sigma_{pp})$ \cite{NuSea:2001idv} & 15 & 16.09 & 13.50 \\
 204 & E866 Drell-Yan process $Q^3d^2\sigma_{pp}/(dQdx_F)$ \cite{NuSea:2003qoe} & 184 & 244.4 & 240.3 \\
 246 & LHCb 8 TeV 2.0 fb$^{-1}$ Z $\rightarrow$ e$^-$e$^+$ forward  & 17 & 25.82 & 23.63 \\
 & rapidity cross section \cite{LHCb:2015kwa} & & & \\
 249 & CMS 8 TeV W cross-section and $A_{\text{ch.}}$ \cite{CMS:2016qqr} & 11 & 11.37 &  8.089 \\ \hline
  & all other data sets & 3141 & 3708 & 3719 \\
  & total & 3681 & 4292 & 4292 \\
 \hline 
 \hline
\end{tabular}
\end{center}
\caption{\label{tb:chi2} The $\chi^2$ of selected data sets included in the CT18 and CT18CS fits, with non-negligible $\Delta \chi^2 = |\chi^2_{\text{CT18}} - \chi^2_{\text{CT18CS}}|$. $N_{pt, E}$ is the number of data points of individual data set, and $\chi^2_{\text{CT18}}$ and 
$\chi^2_{\text{CT18CS}}$ are the $\chi^2$ values obtained by using the central set of CT18 and CT18CS PDFs, respectively. 
}
\end{table}

\subsection{Comparison of PDFs} \label{sub:PDF_shape}

In this section, we compare the fitted CT18CS PDFs obtained in this analysis to the published CT18 PDFs~\cite{Hou:2019efy}.

\begin{figure}[htpb]
	\centering
	\subfigure[]
	{{\includegraphics[width=0.48\hsize]{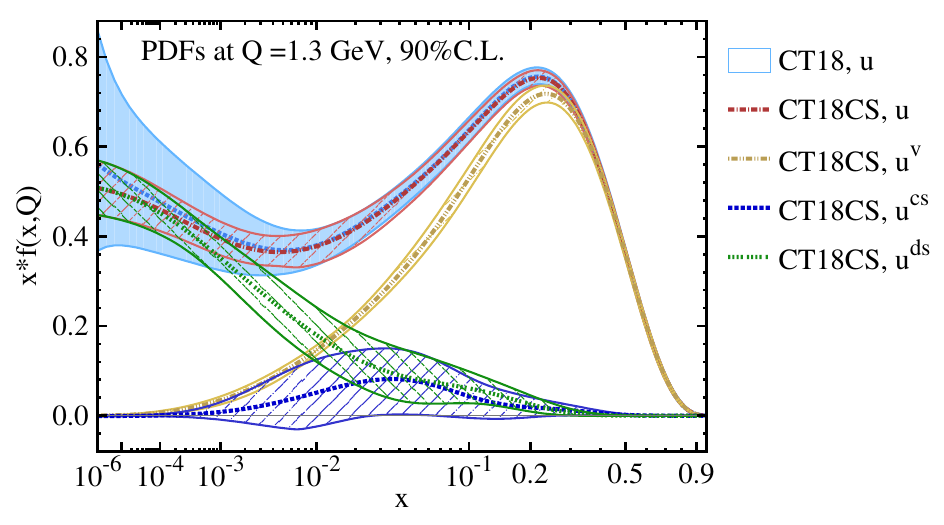}}
		\label{fig:decompose_u_d:a}}
	\subfigure[]
	{{	\includegraphics[width=0.48\hsize]{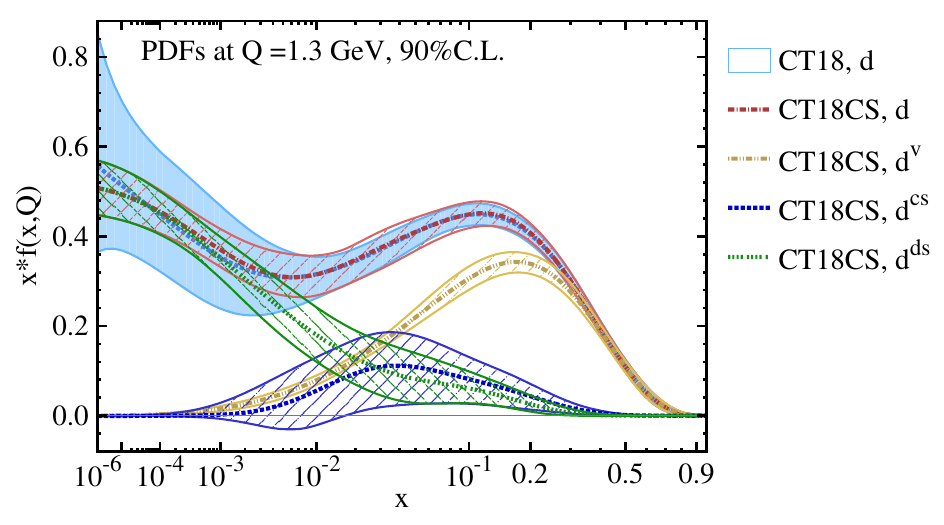}}
		\label{fig:decompose_u_d:b}}

	\caption{\label{fig:decompose_u_d}The decomposition of CT18CS $u$ and $d$ quark distributions in the CS and DS classification at $Q = 1.3$ GeV. The CT18CS PDFs are compared to the nominal CT18 NNLO, which is presented in blue dot line (for central set prediction) and blue error band (for PDF uncertainty).
	}
\end{figure}

\begin{figure}[htpb]  
	\centering
	\subfigure[]
	{{\includegraphics[width=0.48\hsize]{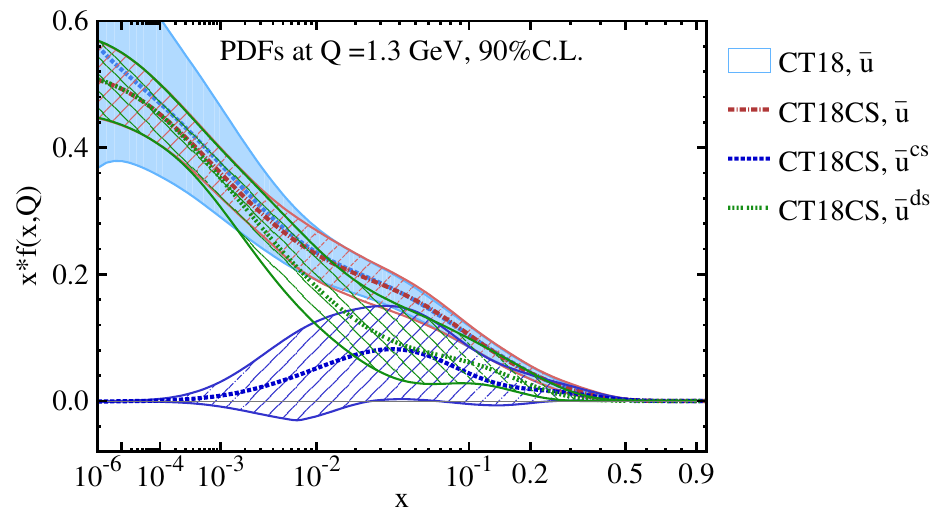}}
		\label{fig:decompose_ubr_dbr_a}}
	\subfigure[]
	{{\includegraphics[width=0.48\hsize]{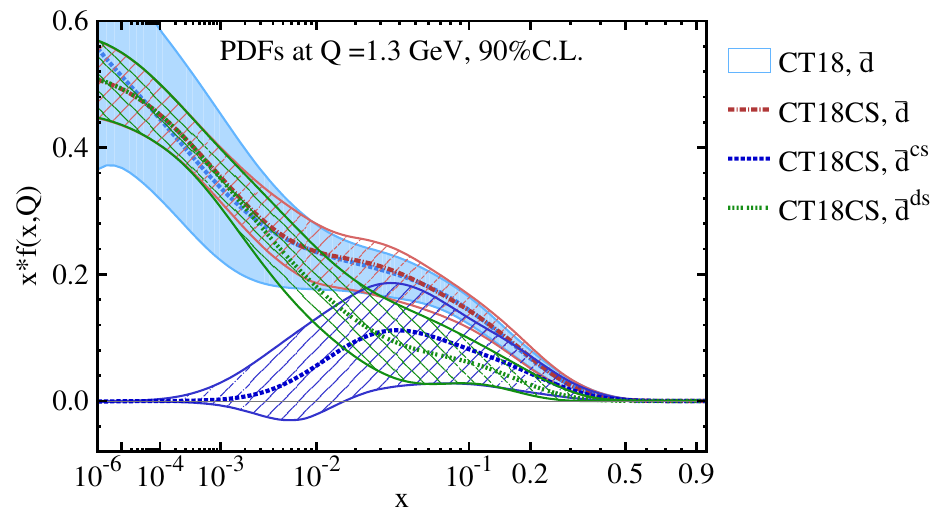}}
		\label{fig:decompose_ubr_dbr_b}}
	\caption{\label{fig:decompose_ubr_dbr}
		Similar to Fig. \ref{fig:decompose_u_d}, but for the decomposition of CT18CS $\bar{u}$ and $\bar{d}$.
	} 
\end{figure}

\begin{figure}[htpb]
	\centering
    \subfigure[]
	{{\includegraphics[width=0.45\hsize]{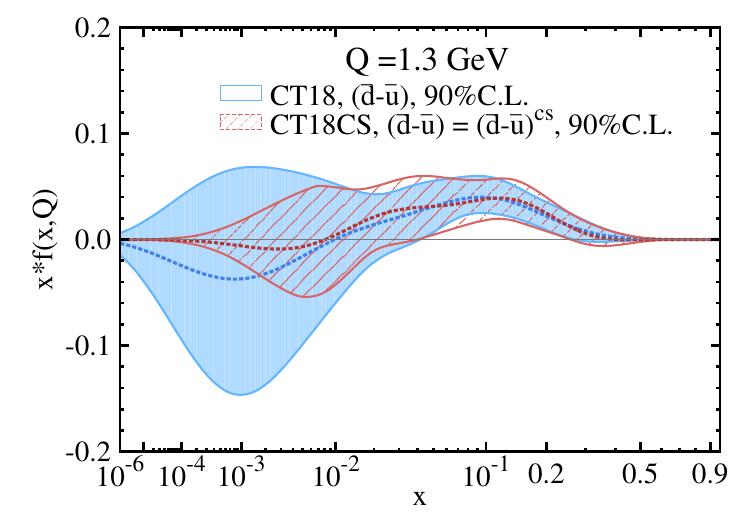}}
	}
	\caption{\label{fig:PDFs_cs}
		Comparison of CT18CS and CT18, for $\bar d - \bar u$. In CT18CS, $(\bar d - \bar u)=(\bar d - \bar u)^{cs}$ due to the ansatz ${\bar d}^{ds}= {\bar u}^{ds}$, cf. Sec.~\ref{sub:PartonDoF}.
	} 
\end{figure}

\begin{figure}[htpb] 
\centering
\subfigure[]
{{\includegraphics[width=0.48\hsize]{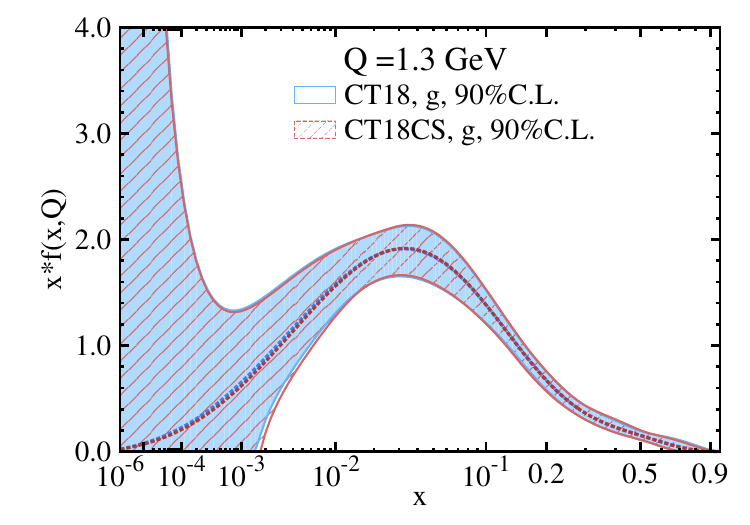}}
 \label{fig:PDFs_g}}
\subfigure[]
{{\includegraphics[width=0.48\hsize]{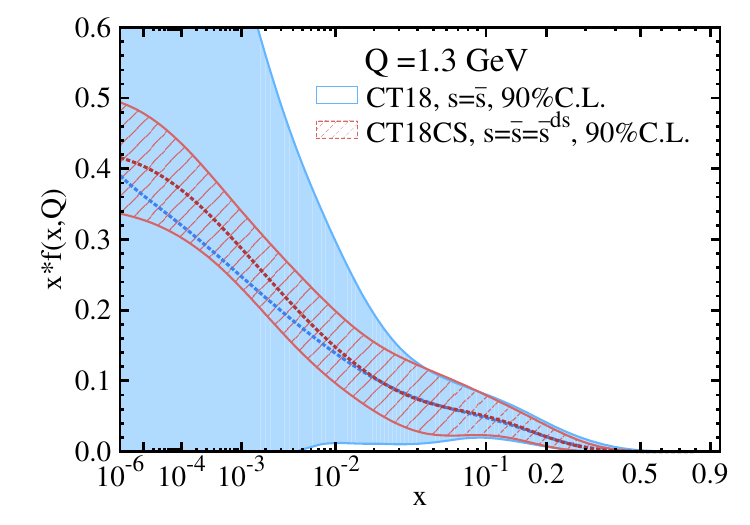}}
 \label{fig:PDFs_ds}}
\subfigure[]
{{\includegraphics[width=0.48\hsize]{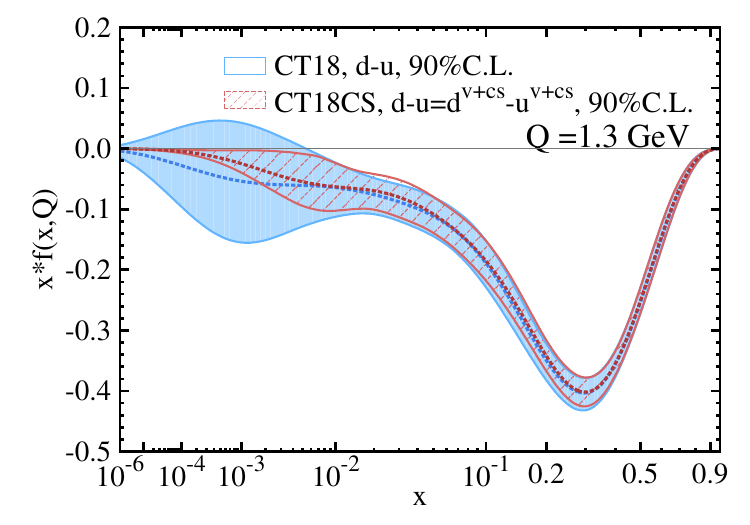}}
 \label{fig:PDFs_d-u}}
\subfigure[]
{{\includegraphics[width=0.48\hsize]{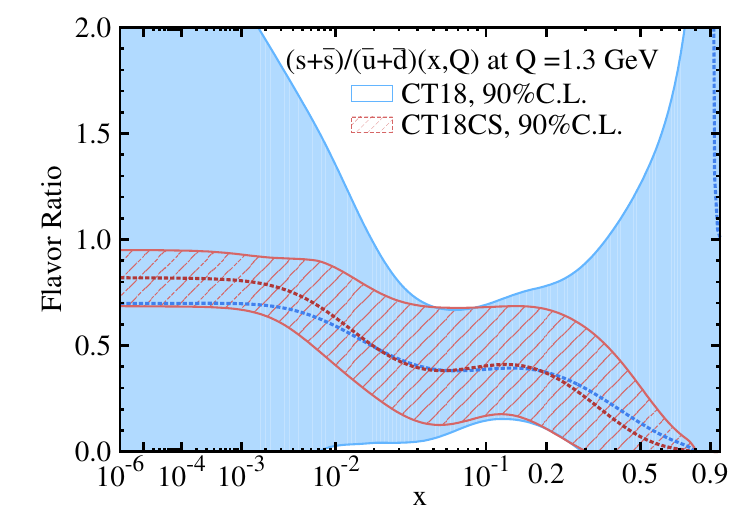}}
 \label{fig:PDFs_Rs}}
\caption{\label{fig:otherPDFs}
Similar to Fig.~\ref{fig:PDFs_cs}, but for comparing $g$-PDF, $s$-PDF, $d-u$ and PDF ratio $(s+{\bar s})/(\bar u + \bar d)$ at $Q=1.3$ GeV.
} 
\end{figure}

In CT18CS, the $u$ and $d$ quark distributions are represented by the combination of valence, connected sea, and disconnected sea quark distributions. For $\bar{u}$ and $\bar{d}$ distributions, they are also made of connected sea, and disconnected sea distributions. In Figs. \ref{fig:decompose_u_d} and \ref{fig:decompose_ubr_dbr}, the decomposition of $u$, $d$, $\bar{u}$, $\bar{d}$ in terms of CS and DS parton distributions, at $Q_0=1.3$ GeV, are shown, respectively. The PDF error bands, obtained at the 90\% confidence level (C.L.), are also shown for comparison.

As shown in Figs. \ref{fig:decompose_u_d:a} and \ref{fig:decompose_u_d:b}, the summation of valence, connected sea, and disconnected sea quark distributions of $u$ and $d$ agrees well with the CT18 central PDF values. The CS components of $u$ and $d$ provide sizable contributions in the intermediate-$x$ region only, i.e., $ 10^{-3} < x < 0.4$.
The $u$ and $d$ PDFs in 
the large-$x$ region are dominated by $u^v$ and $d^v$. At small-$x$, the DS components are, as expected, dominating $u$ and $d$ PDFs, where both the valence and CS components are suppressed. 
Similar comparisons made for 
$\bar{u}$ and $\bar{d}$ are displayed in Fig. \ref{fig:decompose_ubr_dbr}. As shown, CT18CS is in good agreement with CT18 NNLO for these parton distributions. 
Furthermore, in CT18CS, the novel CS parton distribution is found to be responsible for $u$ and $d$ sea quark distributions in the intermediate-$x$ region. 
On the contrary, the DS patron distribution plays a more important role in the small-$x$ region.
We should note that the errors of $u, d, \bar{u}$ and $\bar{d}$ at small $x$ ({\it i.e.,} $x < 10^{-3}$) from CT18CS are substantially smaller than those in CT18. This is mainly due to the ansatz that we imposed on their small $x$ behavior to be $x^{-1}$ in Sec.~\ref{sub:Small-Largex}.

A useful format to compare $\bar d$ and $\bar u$ PDFs resulted from the CT18CS and CT18 fits is to examine their difference, as shown in  
Fig. \ref{fig:PDFs_cs}. Since we have assumed in this analysis that the DS component of $\bar d$ and $\bar u$ are the same, hence $(\bar d - \bar u) = (\bar{d}^{cs} - \bar{u}^{cs})$ in CT18CS PDFs.  
It shows that the CT18CS central value is close to that of CT18 NNLO for $x > 0.03$. In the small-$x$ region, the difference $(\bar{d}^{cs} - \bar{u}^{cs})$ of CT18CS vanishes.
This result is consistent with the prediction presented in Fig. 4 of Ref.~\cite{Liu:2012ch} in which the E866 NuSea~\cite{NuSea:2001idv}  and HERMES~\cite{Airapetian:2008qf} data were compared to CT10 PDFs~\cite{Lai:2010vv} in the framework of leading order analysis. 

We also show in Fig.~\ref{fig:otherPDFs} similar comparison for $g$-PDF, $s$-PDF, $d-u$,  and PDF ratio $(s+{\bar s})/(\bar u + \bar d)$ at $Q=1.3$ GeV.
In Fig. \ref{fig:PDFs_g}, gluon distributions in CT18CS and CT18 fits are in very good agreement across the whole $x$ range. The quark CS and DS separation has no effect on the gluon distribution.
As shown in Fig.~\ref{fig:PDFs_ds}, the uncertainty in the strangeness distribution in CT18CS is reduced by a large margin, as compared to CT18 for $x < 0.03$. This is of the same pattern as for $u,d$ and $\bar{u}, \bar{d}$ since we have adopted the same ansatz, i.e.  $a_1^s=0$, in the CT18CS fit, cf. Sec.~\ref{sub:Small-Largex}. The central value of the $s$-PDF distribution of CT18CS for small $x$ is correlated with those of $\bar{u}$ and $\bar{d}$ via the $R$ ratio introduced from the lattice result. 

The comparison of $d-u$ distribution between CT18CS and CT18 is shown in Fig. \ref{fig:PDFs_d-u}. The $d-u$ distribution corresponds to the $d^{v+cs}-u^{v+cs}$ distribution in CT18CS. This is because $u^{ds}$ and $d^{ds}$ are assumed to be the same under isospin symmetry, cf. Eq.~(\ref{eq:ansatz_equation}). For $x > 0.005$, both central values and sizes of the uncertainty bands of the two PDFs are in good agreement. In the low-$x$ region, the ansatz that CS and valence partons have the same behaviour for $x \rightarrow 0$ in CT18CS leads to a significant reduction in the uncertainty size.

In Fig. \ref{fig:PDFs_Rs}, the ratio of $(s + \bar s)/(\bar u + \bar d)$ in CT18CS is compared to that in CT18. In the small-$x$ region, where the DS parton dominates the sea quark distribution, this ratio for CT18CS is constrained by the lattice input, in addition to the ansatz $a_1^s=0$, which reflects its central value and small uncertainty. As a consequence, the error in CT18CS is greatly reduced as compared to that of CT18 for $x < 0.03$. 
In the larger-$x$ region, where the CS parton becomes important, this ratio is constrained by 
imposing the same large-$x$ behaviors for $\bar{u}$ and $\bar{d}$ as in CT18. 
It is noted~\cite{Liang:2019xdx} that the PDF ratio  
$(s+ \bar{s})/(\bar{u} + \bar{d})$ starts to dip  for $x > 0.01$. This is due to the fact that $\bar{u}$ or $\bar{d}$ has two components -- CS and DS, in contrast to $\bar{s}$ which only has DS. As shown in Fig.~\ref{fig:decompose_ubr_dbr_b}, when $x> 0.01$ the CS components start to show up and contribute to the denominator of the ratio, making it smaller.

\subsection{PDF Mellin Moments} \label{subsec:moments}

The momentum carried by a certain flavour parton can be calculated in terms of the second moment $\langle x \rangle$ of its PDF.  In Table \ref{tb:moment_1}, we compare the 
predictions of CT18CS to CT18 PDFs on the 
second moments of various partons at the input scale.
The $\bar{u}$ and $\bar{d}$ are split into CS and DS in CT18CS and $\langle x\rangle_{u^{v+cs}}$ is from the direct insertion calculation of the $u$ quark on the lattice, which corresponds to the sum of the valence and CS, cf. Eq.~(\ref{valence}).
Other similar comparisons can be found in Table VII of Ref.~\cite{Hou:2019efy}. 
Without the CS and DS separation, one is not able to compare separate flavor-dependent  PDF moments to those from the lattice calculation~\cite{Liu:2017lpe}, since the disconnected insertion lattice calculation corresponds to the DS, while the CS is lumped with the valence in the connected insertion.  The only exceptions are the strange moments which only have DS and 
those of $u - d$ which only involve the connected insertion. They are quite limited. One cannot compare the moments for $u,d, \bar{u}$ and $\bar{d}$.

Now that the CS and DS are separated (although at the input scale) in CT18CS, the lower half of Table \ref{tb:moment_1} shows that, at $Q_0 = 1.3$ GeV, $\bar{u}^{cs}$and $\bar{d}^{cs}$ carry about 1.20\% and 1.97\% of the total momentum of the proton, respectively. 
Namely, $\bar{d}^{cs}$ carries more momentum than $\bar{u}^{cs}$.  
For comparison, $\bar{u}^{ds}$ and $\bar{d}^{ds}$ each carries about 1.67\% of the total momentum of the proton.
Totally, the CS and DS components of up- and down-quarks carry about 6.34\% and 6.68\%  of the total momentum of the proton, respectively. In addition, the strange PDF only has DS component which accounts for 2.74\% of proton's total momentum, with both $s$ and $\bar s$ contributions included.
This is driven by the input value of $R$ taken from the lattice prediction of $1/R = <x>_{s+\bar{s}}/<x>_{\bar{u}+\bar{d}}(\text{DI}) = 0.822(69)(78)$ at $Q=1.3$ GeV, where $<x>_{\bar{u}+\bar{d}}(\text{DI})$ is the momentum fraction carried by the DS component of $\bar{u}$ and $\bar{d}$ partons. 
By separating the CS and DS components of partons in the global analysis, the predictions in Table~\ref{tb:moment_1} can be directly compared to lattice calculations of separate flavors in both the connected and disconnected insertions, term by term. 

In Table \ref{tb:moment_2}, we collect the second moments of $u^+-d^+ = (u+\bar{u})-(d+\bar{d})$ and $s^+=s+\bar{s}$ predicted by CT18CS and CT18 calculations, at 1.3 GeV and 2.0 GeV, respectively. 
Lattice results of $\langle x \rangle_{u^+-d^+}$ and  $\langle x \rangle_{s+}$ at $Q=2.0$ GeV are also given and they are found to be consistent with the CT18 predictions. However, we note that the deviation of the lattice calculations from
different groups are large and not all systematic errors have been taken into account.

\begin{table}[htpb]
\begin{center}
\begin{tabular}{l|rrrrrrr}
      PDF     & $<x>_{u^{v}}$ & $ <x>_{d^{v}}$   & $<x>_{g}$  & $<x>_{\ub}$ &  $<x>_{\db}$  & $<x>_{s}$           \\
  \hline
  CT18  & $0.325(5)$  & $0.134(4)$  & $0.385(10)$  & $0.0284(22)$ &  $0.0361(27)$  & $0.0134(52)$    \\
    CT18CS  & $0.323(4)$ & $0.136(3)$ & $0.384(12)$ & $0.0287(25)$ & $0.0364(34)$  & $0.0137(39)$  \\ \hline \hline
    PDF     & $<x>_{u^{v+cs}}$ & $<x>_{d^{v+cs}}$  & $<x>_{\ub^{cs}}^*$ & $<x>_{\db^{cs}}^*$ & $<x>_{u^{ds}}^{\dagger}$  \\
  \hline                                                    CT18CS  &  $0.335(7)$ & $0.155(8)$ & $0.0120(64)$ & $0.0197(70)$ & $0.0167(49)$ \\
\end{tabular}
\end{center}
\caption{\label{tb:moment_1} The second moment $\langle x \rangle$  of CT18CS and CT18 NNLO at $1.3$ GeV. The superscript ``$*$" indicates that due to the fourth ansatz imposed in Eq. (\ref{eq:cs_ansatz}), the second moments for CS components between quarks and anti-quarks are identical, namely, $<x>_{\ub^{cs}} = <x>_{u^{cs}}$ and $<x>_{\db^{cs}} = <x>_{d^{cs}}$. The superscript  ``$\dagger$" indicates that due to the second ansatz imposed in Eq. (\ref{eq:ansatz_equation}), the second moments of DS components of $u$, $\bar{u}$, $d$, and $\bar{d}$ are identical.
}
\end{table}

\begin{table}[htpb]
\begin{center}
	\begin{tabu}{l|ll|ll}
 & $Q = 2.0$ GeV & & $Q= 1.3$ GeV & \\
                                    & \text{CT18} & \text{Lattice} & \text{CT18CS} & \text{CT18}\\
\hline

                                      &             & 
                                      $0.111\!-\!0.209^{N_f=2+1}$            &       &      \\
 $\langle x \rangle_{u^{+} - d^{+}}$  & $0.156(7)$    & $0.153\!-\!0.194^{N_f=2+1+1}$~\cite{Lin:2020rut}    & $0.173(7)$ & $0.175(8)$ \\
                                      &             & $0.166\!-\!0.212^{N_f=2}$              &       &      \\
 $\langle x \rangle_{s^{+} }$         & $0.033(9)$    & $0.051(26)(5)$ \cite{Yang:2018nqn}     & $0.027(8)$  & $0.027(10)$ \\
\end{tabu}
\end{center}
\caption{\label{tb:moment_2} The second moments of $(u^+-d^+)$ and $s^+$ predicted by CT18~\cite{Hou:2019efy} and CT18CS at 2.0 GeV and 1.3 GeV, respectively. We also show lattice results at 2.0 GeV.
For $\langle x \rangle_{u^+-d^+}$, we follow
Ref.~\cite{Lin:2020rut} in supplying ranges obtained from various calculations, grouped according to the number
of active flavours, $N_f$, in the lattice action used.
}
\end{table}

\section{The Impact of SeaQuest data} \label{E866E906} 

\begin{figure}[htpb]  
	\centering
	\subfigure[]
	{{\includegraphics[width=0.45\hsize]{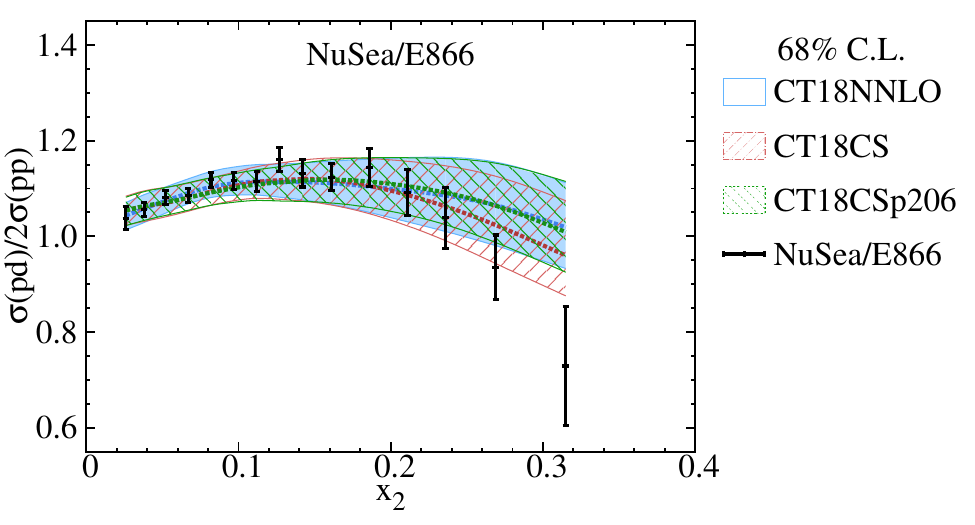}}
		\label{fig:abs_data-theory_E866}}
	\subfigure[]
	{{\includegraphics[width=0.45\hsize]{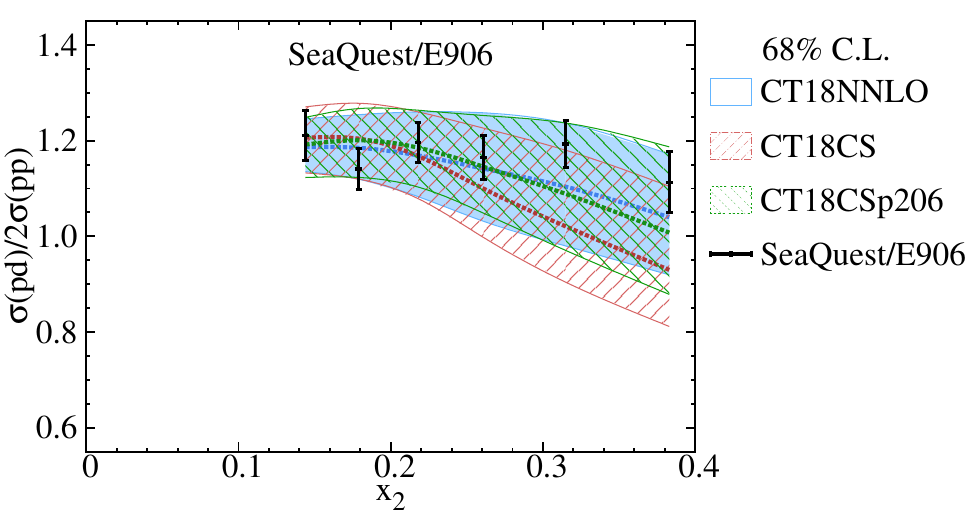}}
		\label{fig:abs_data-theory_E906}}
	\caption{\label{fig:abs_E866_E906} 
	Comparison of CT18 NNLO, CT18CS and {CT18CSp206} predictions to the (a) E866 SeaQuest and (b) E906 SeaQuest data. Note that SeaQuest data were not included in the CT18NNLO and CT18CS fits, but are in the {CT18CSp206} fit.
	} 
\end{figure}

Fixed-target Drell-Yan measurements provide an important probe of the $x$ dependence of the nucleon PDFs. This fact motivated the Fermilab E866 NuSea experiment \cite{NuSea:2001idv}, which determined the deuteron-to-proton cross section ratio
$\sigma_{pd} \big/ 2\sigma_{pp}$ out to relatively large $x_2$, the momentum fraction of the target. 
Intriguingly, E866 found evidence that the cross section ratio dropped below unity, $\sigma_{pd} \big/ 2\sigma_{pp} < 1$,
as $x_2$ approached and exceeded $x \gtrsim 0.25$. 
The E866 results stimulated an interest in performing a similar measurement out to larger $x_2$ with higher
precision --- the main objective of the subsequent  E906 SeaQuest experiment at Fermilab \cite{SeaQuest:2021zxb}.
Comparing to the NuSea data, the recent SeaQuest data include an extra bin which records data around $x \sim 0.4$ with high precision.
In Fig. \ref{fig:abs_E866_E906}, we compare the predictions by CT18CS to the NuSea and SeaQuest data. For $x_2 > 0.2$, the NuSea and the SeaQuest data exhibit different shapes of $\sigma(pd)/2\sigma(pp)$. The ratio $\sigma(pd)/2\sigma(pp)$ for the NuSea data clearly decreases as $x_2$ becomes higher than 0.2, while for the SeaQuest data, this ratio seems to remain the same up to $x_2=0.4$. The difference in the shape of $\sigma(pd)/2\sigma(pp)$ distribution implies that NuSea and SeaQuest data have different preference for the PDF-ratio ${\bar d}/{\bar u}$ or the PDF-difference ${\bar d} - {\bar u}$ in the large-$x$ region. 
In view of the fact that, in the CT18CS analysis, ${\bar q}={\bar q}^{cs}+{\bar q}^{ds}$ for $q=u$ or $d$ and ${\bar u}^{ds}={\bar d}^{ds}$, the deviation of ${\bar d}/{\bar u}$ from unity is thus due to the different ${\bar u}^{cs}$ and ${\bar d}^{cs}$ contributions in the proton.
Hence, it is interesting to know how the inclusion of the SeaQuest data in a global fit such as CT18CS could modify the PDF-difference $(\bar d - \bar u)$, cf.  Fig.~\ref{fig:PDFs_cs}, which is equal to $(\bar d - \bar u)^{cs}$.
\begin{table}[htpb]
	\begin{center}
		\begin{tabular}{cl|r|rr}
			ID & Experimental data set & $N_{pt, E}$ & $\chi^2_{\text{CT18CS}}$ & $\chi^2_{\text{CT18CSp206}}$ \\ \hline \hline
			203 & E866 Drell-Yan process $\sigma_{pd}/(2\sigma_{pp})$ \cite{NuSea:2001idv} & 15  & 13.5 & 18.8 \\
			\hline 
			\hline
			206 & E906 SeaQuest Drell-Yan process $\sigma_{pd}/(2\sigma_{pp})$ \cite{SeaQuest:2021zxb} & 6 & 20.6 & 8.24 \\
		\end{tabular}
	\end{center}
	\caption{\label{tb:chi2new} The $\chi^2$ of selected data sets included in the CT18CS and CT18CSp206 fits. Only those with non-negligible $\Delta \chi^2 = |\chi^2_{\text{CT18CS}} - \chi^2_{\text{CT18CSp206}}|$ are listed. $N_{pt, E}$ is the number of data points of individual data set, and $\chi^2_{\text{CT18CS}}$ and $\chi^2_{\text{CT18CSp206}}$ are the $\chi^2$ values predicted by using the CT18CS and the CT18CSp206 fit.	
	Note that the E906 SeaQuest data \cite{SeaQuest:2021zxb} (ID=206) are not included in the CT18CS fit, but are in the CT18CSp206 fit.
}
\end{table}

\begin{figure}[htpb]  
	\centering
	\subfigure[]
	{{\includegraphics[width=0.45\hsize]{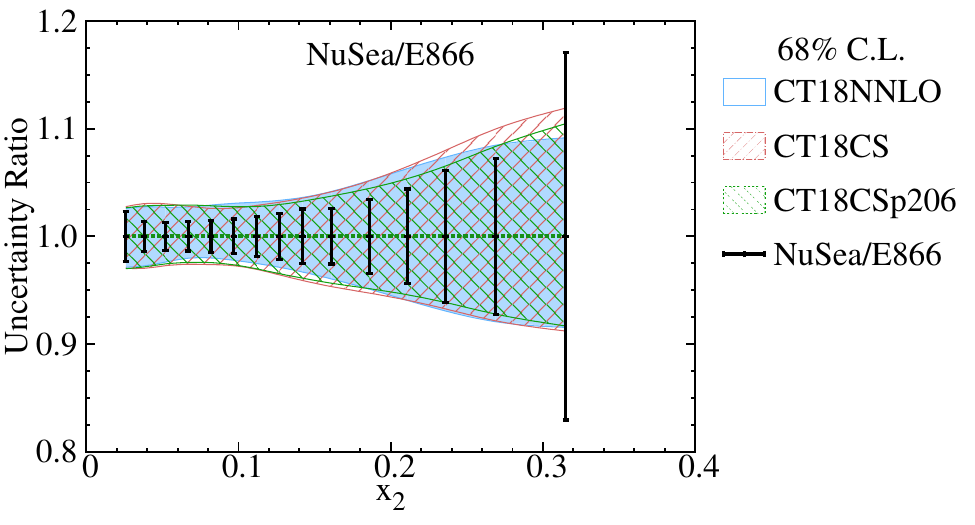}}
		\label{fig:unc_data-theory_E866}}
	\subfigure[]
	{{\includegraphics[width=0.45\hsize]{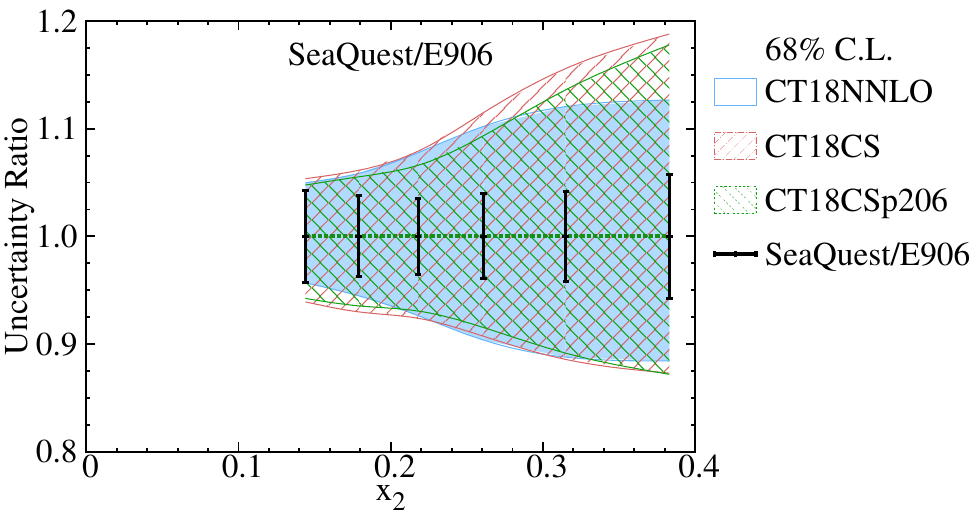}}
		\label{fig:unc_data-theory_E906}}
	\caption{\label{fig:unc_E866_E906} 
	Similar to Fig. \ref{fig:abs_E866_E906}, but for the comparison of the sizes of (total) experimental uncertainty of (a) E866 NuSea and (b) E906 SeaQuest experiments to the PDF-induced uncertainty predicted by CT18NNLO, CT18CS and CT18CSp206. 
	} 
\end{figure}

Below, we discuss the result of a new fit, referred to as ``CT18CSp206'' below, which follows the same approach as CT18CS, but with the inclusion of the E906 SeaQuest data to the original CT18 data set. 
In Table~\ref{tb:chi2new}, we compare the quality of the CT18CSp206 fit to that of CT18CS.  
The only data sets with non-negligible $\Delta \chi^2 = |\chi^2_{\text{CT18CS}} - \chi^2_{\text{CT18CSp206}}|$ are just the E866 NuSea data and E906 SeaQuest data. From CT18CS to {CT18CSp206} fit, the $\chi^2$ for E866 NuSea data is increased by about 5 units, while the fit to the E906 SeaQuest data is improved (with a reduction of 12 units in its $\chi^2$). This tension in the change of $\chi^2$ reflects the different preferences of PDF-ratio $\bar d/\bar u$ or the PDF-difference $\bar d - \bar u$ in the large-$x$ region.
In Figs. \ref{fig:abs_E866_E906} and  \ref{fig:unc_E866_E906}, we compare the predictions of CT18CS and {CT18CSp206} to the NuSea and SeaQuest data. 
In Fig. \ref{fig:abs_data-theory_E866}, the prediction of CT18CS is closer to the E866 NuSea data points for $x_2 > 0.2$, comparing to those of CT18 and {CT18CSp206}. For Fig. \ref{fig:abs_data-theory_E906}, the CT18CS prediction presents a different shape from E906 SeaQuest data points particularly for $x_2 > 0.2$, while CT18 and {CT18CSp206} PDFs show better consistencies with these data points.
Fig. \ref{fig:unc_E866_E906} shows the comparison of uncertainty sizes between the total experimental uncertainty and the PDF-induced uncertainty in predictions for both E866 NuSea and E906 SeaQuest data. All of three above-mentioned PDFs sets exhibit conservative uncertainties, so that the PDF-induced uncertainties in predictions are larger than the experimental uncertainty for both data sets, except for the data point with the highest $x_2$ value in E866 NuSea measurement.
For $x_2 > 0.2$, the CT18CS predictions for both data sets possess a slightly larger error bands than predictions of CT18 and {CT18CSp206}. For most of the range of $x_2$, the error band of  {CT18CSp206} is comparable to the CT18 error band, while in the prediction of E906 SeaQuest data with $x_2 > 0.3$, {CT18CSp206} has a larger uncertainty.

\begin{figure}[htpb]  
	\subfigure[]
	{{\includegraphics[width=0.56\textwidth]{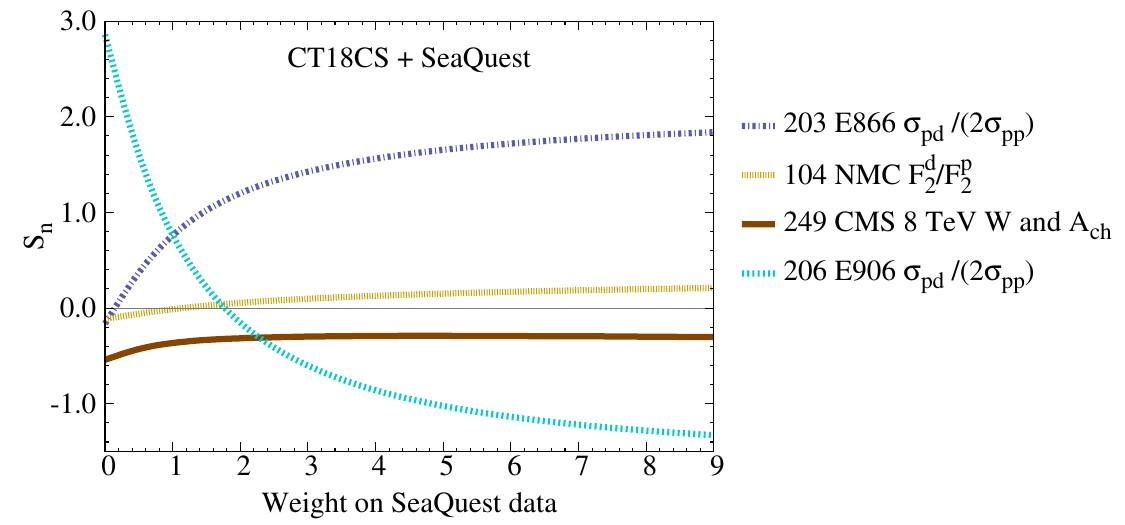}} \label{fig:delta_SE}}	
	\subfigure[]
	{{\includegraphics[width=0.38\textwidth]{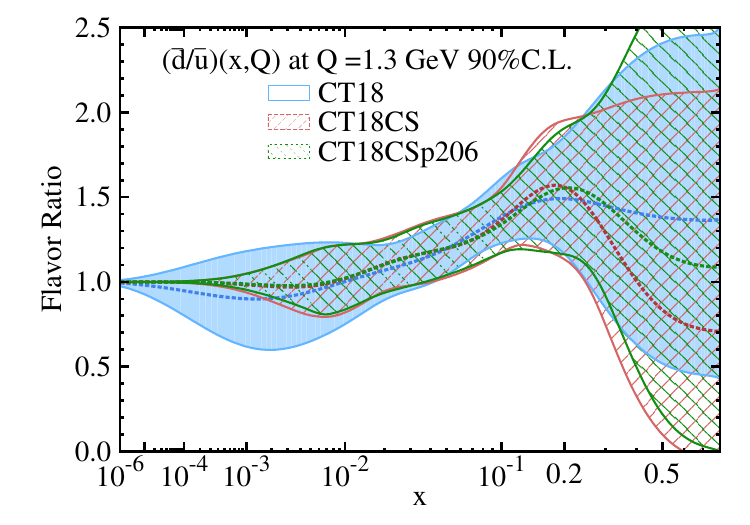}} \label{fig:dbarOubar_CT18}}
	\caption{\label{fig:epump}
		(a) The change of the effective Gaussian variable $S_E$ for some data sets included in CT18CS, as the weight of E906 SeaQuest data increases from 0 to 10. Only the data sets with notable changes in $S_E$ are shown. Note that a weight of zero corresponds to the CT18CS fit, in which the SeaQuest data is not included.
		(b) Comparison of the PDF ratio $\bar{d}/\bar{u}$, as a function of $x$ at $Q=1.3$ GeV, among CT18, CT18CS, and  {CT18CSp206}.
}
\end{figure}

\begin{figure}[htpb]  
		\subfigure[]
	{{\includegraphics[width=0.45\textwidth]{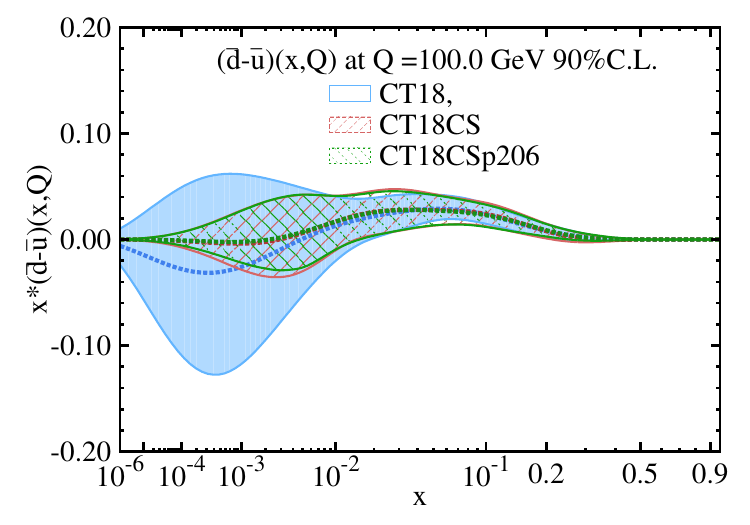}} \label{fig:dbr_ubr_100GeV:a}}
		\subfigure[]
	{{\includegraphics[width=0.45\textwidth]{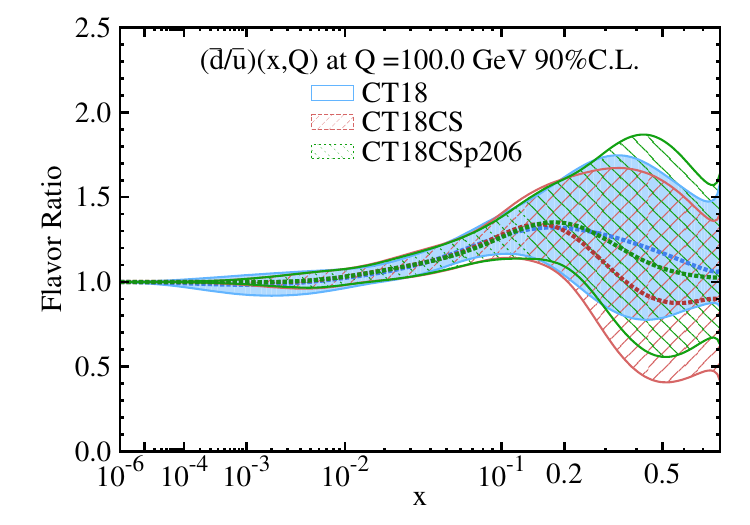}} \label{fig:dbr_ubr_100GeV:b}}  \\
		\subfigure[]
	{{\includegraphics[width=0.45\textwidth]{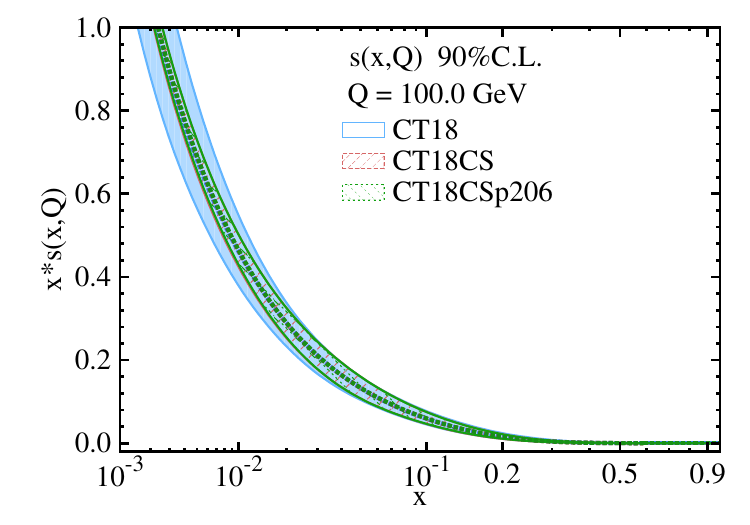}} \label{fig:s_100GeV:c}}		
		\subfigure[]
	{{\includegraphics[width=0.45\textwidth]{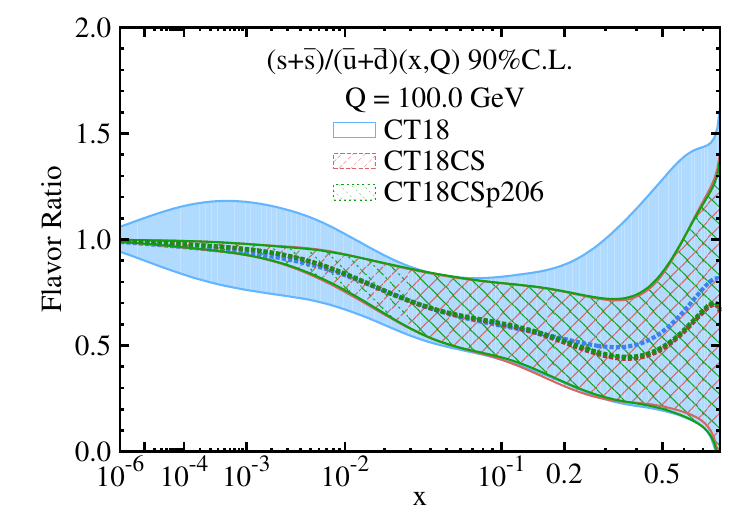}} \label{fig:rs_100GeV:d}}		
	\caption{\label{fig:dbr_ubr_100GeV}
	The comparison of $\bar{d} - \bar{u}$, $\bar{d}/\bar{u}$, $s(x)$, and $(s+\bar{s})/(\bar{u}+\bar{d})$ PDFs at 100 GeV for for CT18, CT18CS, and CT18CSp206.
		}
\end{figure}

Finally, we remark that the impact of SeaQuest data to modifying the CT18CS PDFs can also be studied by using the  ePump-updating method, detailed in Refs.~\cite{Schmidt:2018hvu,Hou:2019gfw}. 
The idea is to add the SeaQuest data, with a given weight, to the original CT18 data set and perform a new global fit using the ePump-updating method. This will update the original CT18CS  PDFs and produce a new set of PDFs. Given this new set of PDFs, one can calculate the change in the total $\chi^2$ of each data set included in the global fit, as compared to that given by the original CT18CS PDFs. 
Instead of examining $\chi^2_E(N_{pt,E})/N_{pt,E}$ for the  individual experiment $E$, which has different probability distribution and is dependent on the total number of data point $N_{pt,E}$, we provide an equivalent information in the form of the effective Gaussian variables $S_E=\sqrt{2\chi^2_E}-\sqrt{2N_{pt,E}-1}$~\cite{Lai:2010vv}.
A well-fitted data set should have $S_E$ between $-1$ and 1. An $S_E$ smaller than $-1$ means the
data set is fitted too well (maybe due to large experimental errors) and an $S_E$ larger than 1 indicates poor fitting.
To examine the potential tensions between the E906 SeaQuest data and the data sets included in the CT18CS fit, we plot 
in Fig.~\ref{fig:delta_SE}
the change of the effective Gaussian variable $S_E$ for some data sets included in CT18CS as the weight of SeaQuest data is increased from 0 to 10. Only the data sets with non-negligible  change in $S_E$ are shown. Note that a weight of zero corresponds to the CT18CS fit, in which the SeaQuest data were not included, and a weight of one leads to the above-mentioned CT18CSp206 fit.
As the weight of SeaQuest data increases, the $S_E$ of 
SeaQuest data becomes smaller, as expected, while the E866 NuSea data becomes much larger, indicating 
tension with the SeaQuest data. 
Both the NMC $F_2^d/F_2^p$ (ID=104) and CMS 8 TeV $W$ and $A_{ch}$ (ID=249) data show very slight increase in their $S_E$ values as the weight of SeaQuest data increases from zero.
In Fig.~\ref{fig:dbarOubar_CT18}, we compare the PDF-ratio  $\bar{d}/ \bar{u}$, as a function of $x$ at $Q=1.3$ GeV, 
among CT18 NNLO, CT18CS, and  {CT18CSp206}, where the E906 SeaQuest data (labelled as ID=206 in Table~\ref{tb:chi2new}) is included, via the ePump-updating method~\cite{Schmidt:2018hvu,Hou:2019gfw}. It shows that {CT18CSp206} has a larger PDF ratio $\bar d/\bar u$ at $x > 0.2$, as  compared to CT18CS. On the other hand, the uncertainty of the PDF ratio $\bar d/\bar u$ of {CT18CSp206} in large-$x$ region is enlarged from that of CT18CS to tolerate the tension between the two data sets.
For completeness, we also show in Fig.~\ref{fig:dbr_ubr_100GeV} the comparison of $\bar d/\bar u$, $(\bar d - \bar u)$, $s$, and $(s+{\bar s})/({\bar u}+{\bar d})$, respectively, as predicted by these three different global fits at $Q=100 $ GeV. In Figs.~\ref{fig:dbr_ubr_100GeV:a}, \ref{fig:dbr_ubr_100GeV:b}, at 100 GeV the comparison of PDF ratio $\bar d/\bar u$, or of the PDF difference $(\bar d - \bar u)$, is similar to those at 1.3 GeV, c.f. Fig.~\ref{fig:dbr_ubr_100GeV}. The impact of the SeaQuest data on $s(x)$ and PDF ratio $(s+{\bar s})/({\bar u}+{\bar d})$ at 100 GeV is negligible, as shown in Figs.~\ref{fig:s_100GeV:c}, \ref{fig:rs_100GeV:d}.

\section{Summary} \label{summary} 

In this work, we present a NNLO QCD global analysis named CT18CS where the connected sea partons and disconnected sea partons, as revealed in the path-integral formulation of the hadronic tensor in QCD, are separately parametrized at the input scale of $Q_0 = 1.3$ GeV.
The CS and DS are mainly distinguished by their respective small-$x$ behaviors. Furthermore, we assumed that the DS of $u$ and $d$ are proportional to the $s$ with the proportional constant constrained by a recent complete lattice calculation of the second moment ratio~\cite{Liang:2019xdx} $\langle x \rangle_{s+\bar{s}}/\langle x \rangle_{\bar{u}+\bar{d}}(\text{DI}) = 0.822(69)(78)$ at $Q=1.3$ GeV, where $\langle x \rangle_{\bar{u}+\bar{d}}(\text{DI})$ is the momentum fraction carried by the DS component of $\bar{u}$ and $\bar{d}$ 
partons. This lattice QCD constraint was included in the CT18CS fit via the Lagrange multiplier method.
Together with the ansatz $a_1^s =0$, this lattice input has helped reduce the error of the ratio $\langle x \rangle_{s+\bar{s}}/\langle x \rangle_{\bar{u}+\bar{d}}$ greatly for $x< 0.03$ as compared to that of the CT18 fit.

Short of applying the evolution equations where CS and DS partons are evolved separately, we impose a number of ansatzes regading small-$x$ behaviors and isospin symmetry, as described in Sec. \ref{sub:PartonDoF}.
in the input scale and evole the combined CS and DS partons during evolution. In this way, the PDFs are still evolved from  $Q_0$ with the usual parton classification, namely $g, u, \bar{u}, d, \bar{d}, s$ and the results can be compared with CT18 at $Q_0$.

It is found that the fit quality of CT18CS is comparable to that of CT18 NNLO. The CT18CS PDFs,
obtained with an extended parametrisation, are consistent with CT18 NNLO in a wide range of $x$, but the errors of the quark partons in CT18CS are greatly reduced at small 
$x$ as compared to those of CT18, mainly due to the small-$x$ behaviors imposed and the lattice QCD input.  As expected, the DS components primarily contribute to $\bar{u}$ and $\bar{d}$ in small-$x$ region, and the CS components provide sizable contribution in the intermediate-$x$ region.
We give the second moments of CS and DS in different flavors at scale $Q_0$. They can be compared with systematic error controlled lattice calculations term by term for the first time. At $Q = 1.3$ GeV, we find that up and down quarks in the CS sector takes about 6.34\% of total momentum, while the momentum in DS sector is about 6.68\% of total amount. They are comparable in size at this low scale.
The implication of CT18CS PDFs are studied in the comparison of predictions for the NuSea data and SeaQuest data between CT18CS and CT18 NNLO PDFs.
A new global fit (referred to as CT18CSp206) on the basis of CT18CS is obtained with the SeaQuest data included. Through a scan of of the effective Gaussian variable $S_E$ over various weights to the E906 SeaQuest data, 
using the ePump-updating method~\cite{Schmidt:2018hvu,Hou:2019gfw},
it is found that the SeaQuest data and the NuSea data are in tension.

In the future, global analyses should incorporate the extended evolution equations~\cite{Liu:2017lpe} where the connected sea and the the disconnected sea are evolved separately so that they will remain
separated at all $Q^2$ for better and more detailed delineation of the PDF degrees of freedom and compared to lattice results term by term.

\section*{Acknowledgment}

The authors are indebted to  J.C. Peng, J.W. Qiu, and Y.B. Yang for insightful discussions.  The work of K.-F. Liu is partially support by the U.S. DOE grant DE-SC0013065 and DOE Grant No.\ DE-AC05-06OR23177 which is within the framework of the TMD Topical Collaboration.
This research used resources of the Oak Ridge Leadership Computing Facility at the Oak Ridge National Laboratory, which is supported by the Office of Science of the U.S. Department of Energy under Contract No.\ DE-AC05-00OR22725. This work used Stampede time under the Extreme Science and Engineering Discovery Environment (XSEDE), which is supported by National Science Foundation Grant No. ACI-1053575.
We also thank the National Energy Research Scientific Computing Center (NERSC) for providing HPC resources that have contributed to the research results reported within this paper.
We acknowledge the facilities of the USQCD Collaboration used for this research in part, which are funded by the Office of Science of 
the U.S. Department of Energy.
The work of J. Liang is partially supported by the National Science Foundation of China (NSFC) under Grant No. 12175073.
The work of C.-P. Yuan is partially supported by the 
U.S.~National Science Foundation
under Grant No.~PHY-2013791. C.-P.~Yuan is also grateful for the support from the Wu-Ki Tung endowed chair in particle physics.

\bibliographystyle{unsrt}
\bibliography{bibliography}

\end{document}